\documentclass[aps,prb,twocolumn,amsmath,amssymb,floatfix,epsfig]{revtex4}
\usepackage{graphicx,epsfig,psfrag,subfigure}
\usepackage{color}
\usepackage{hyperref}
\usepackage{hypcap}
\usepackage{amssymb,amsbsy,amsmath}
\newcommand\beq{\begin{equation}}
\newcommand\eeq{\end{equation}}
\newcommand\bea{\begin{eqnarray}}
\newcommand\eea{\end{eqnarray}}
\newcommand\al{\alpha}
\newcommand\be{\beta}
\newcommand\ga{\gamma}
\newcommand\de{\delta}
\newcommand\ep{\epsilon}
\newcommand\De{\Delta}

\newcommand\om{\omega}
\newcommand\ta{\theta}
\newcommand\dg{\dagger}

\newcommand\non{\nonumber}
\newcommand\noi{\noindent}

\newcommand\ig{\includegraphics}
\newcommand\bib{\bibitem}

\begin{document}

\title{Generating end modes in a superconducting wire by periodic driving 
of the hopping} 

\author{Siddhartha Saha$^1$, Shankar N. Sivarajan$^1$, and Diptiman Sen$^2$} 
\affiliation{\small{
$^1$Undergraduate Programme, Indian Institute of Science, Bengaluru
560 012, India \\
$^2$Centre for High Energy Physics, Indian Institute of Science, Bengaluru
560 012, India}}

\date{\today}

\begin{abstract}
We show that harmonic driving of either the magnitude or the phase of the 
nearest-neighbor hopping amplitude in a $p$-wave superconducting wire can 
generate modes localized near the ends of the wire. The Floquet eigenvalues 
of these modes can either be equal to $\pm 1$ (which is known to occur in 
other models) or can lie near other values in complex conjugate pairs which 
is unusual; we call the latter anomalous end modes. All the end modes have 
equal probabilities of particles and holes. If the amplitude of driving is 
small, we observe an interesting bulk-boundary correspondence for the 
anomalous end modes: the Floquet eigenvalues and the peaks of the Fourier 
transform of these end modes lie close to the Floquet eigenvalues and momenta 
at which the Floquet eigenvalues of the bulk system have extrema. 
\end{abstract}

\maketitle

\section{Introduction}
\label{sec:int}



The last several years have witnessed extensive studies of topological phases 
of matter~\cite{hasan,qi,fidkowski1}. A system in a topological phase has 
only gapped states in the bulk but ha gapless states at the boundaries. 
In addition, the number of gapless boundary modes is given by a topological 
invariant which depends on the properties of the bulk and its symmetries 
such as time-reversal and particle-hole symmetry. Such a relation between
the properties of the bulk and the boundary modes is called a bulk-boundary
correspondence.

Recently, there have been several studies of systems in which the Hamiltonian 
is varied in time in a periodic way leading to some topological features such 
as the generation of boundary modes~\cite{kita1,lind1,jiang,gu,kita2,lind2,
trif,russo,basti1,liu,tong,cayssol,thakur1,rudner,basti2,tomka,gomez,dora,
katan,kundu,basti3,schmidt,reynoso,wu,perez1,perez2,perez3,reichl,thakur2,
asboth,roy,thakur3}. Some of these topological aspects have been 
experimentally studied~\cite{kita3,rechta,rechtb,rechtc,tarruell,jotzu}. 
However, the existence of topological 
invariants and the relation between them and the number of boundary modes is 
not always clear. Further, in many models, the boundary modes turn out to
be of only two types corresponding to eigenvalues of the Floquet operator 
being $+1$ or $-1$. It would be interesting to know if this is always 
the case. In this paper, we examine some of these questions for a 
one-dimensional model where the end modes and topological invariants 
can be numerically studied relatively easily. 

The plan of this paper is as follows. In Sec.~\ref{sec:kit} we introduce 
the system of interest. We will consider a lattice model of spinless electrons
with $p$-wave superconducting pairing; this is
sometimes called the Kitaev chain~\cite{kitaev}. We will review the energy 
spectrum and the different phases (topological and non-topological)
that this model has when the Hamiltonian is time-independent. The phase 
diagram is known to change in an interesting way if the hopping amplitude 
is allowed to be complex; we get regions in parameter space where the bulk 
spectrum is gapless~\cite{gott2}. We present two topological invariants 
which one-dimensional models with and without time-reversal symmetry have when 
periodic boundary conditions are imposed. In Sec.~\ref{sec:flo}, 
we discuss in general how we can numerically study the Floquet time evolution 
and the modes which appear at the ends of a system when the Hamiltonian 
varies periodically with time. In Sec.~\ref{sec:end}, we study what happens 
when the phase or magnitude of the hopping is driven harmonically in time. 
We study the ranges of parameters in which modes appear at the ends of an 
open chain and various properties of these modes such as their number and 
Floquet eigenvalues. We find that the Floquet eigenvalues can be equal to 
either $\pm 1$ or any other complex number with unit magnitude; in the latter 
case they have to appear in complex conjugate pairs, and we call these 
anomalous end modes. We calculate the Fourier transforms of the wave functions
of the end modes and find that they have peaks at certain values of $k$;
in particular, the Fourier transforms of the anomalous end modes have peaks 
at zero and $\pi$. The expectation value of the electron number is found to be
zero in all the end modes; hence they have equal probabilities of particles 
and holes. We find that the anomalous end modes disappear when the 
chemical potential is moved sufficiently away from zero. In 
Sec.~\ref{sec:bbc} we examine if there are any bulk-boundary correspondences in 
this periodically driven system. We first examine a topological invariant
called the winding number and find that it matches the 
number of modes at each end of the chain which have Floquet eigenvalues
equal to 1 for the case that the magnitude of the hopping is
periodically driven but not in the case that the phase of the hopping is 
driven. A corresponding topological invariant for the anomalous
end modes does not seem to exist. When the amplitude of the periodic driving 
is small, we find a different kind of bulk-boundary correspondence 
which works for all the anomalous end modes. Namely, if we look at the 
Floquet eigenvalues of the bulk system as a function of the momentum $k$,
we find that the values of $k$ where these eigenvalues have extrema and the 
corresponding Floquet eigenvalues match closely the values of $k$ where the 
Fourier transforms of the wave functions of the anomalous end modes have peaks
and the Floquet eigenvalues of those end modes. In Sec.~\ref{sec:mag}, we 
use a Floquet-Magnus expansion to study the system when the driving 
frequency is much larger than the other energy scales like the hopping
and the superconducting pairing. In this limit, we find 
that the number of end modes is the same as that found when there is no 
driving. We summarize our main results and point out possible directions for 
future studies in Sec.~\ref{sec:con}.

\section{Kitaev chain}
\label{sec:kit}

In this section, we will review the properties of the Kitaev chain, its 
phase diagram, and topological invariants. The Kitaev chain is a model of 
spinless electrons on a lattice with a nearest-neighbor hopping amplitude 
$\ga$, a $p$-wave superconducting 
pairing $\De$ between neighboring sites, and a chemical potential $\mu$. 
For a finite and open chain with $N$ sites, the Hamiltonian takes the form
\bea H &=& \sum_{n=1}^{N-1} [ \ga f_n^\dg f_{n+1} + \ga^* f_{n+1}^\dg f_n 
+ \De ( f_n f_{n+1} + f_{n+1}^\dg f_n^\dg )] \non \\
&& - \sum_{n=1}^N \mu f_n^\dg f_n, \label{ham1} \eea
where $\De$ and $\mu$ are real, but $\ga$ may be complex. We write the hopping
as 
\beq \ga ~=~ \ga_0 ~e^{i \phi}, \eeq
where $\ga_0$ is real and positive. We will assume that all these parameters 
are time-independent in this section. The operators $f_n$ in Eq.~\eqref{ham1} 
satisfy the anticommutation relations $\{ f_m, f_n \} =0$ and $\{ f_m, f_n^\dg
\} = \de_{mn}$. (We will set both Planck's constant $\hbar$ and the lattice 
spacing equal to 1 in this paper). We introduce the Majorana operators 
\beq b_{2n-1} = f_n + f_n^\dg ~~~~{\rm and}~~~~ b_{2n} = i (f_n - f_n^\dg),
\label{majo} \eeq
for $n=1,2,\cdots,N$. It is easy to check that these
are Hermitian operators satisfying $\{ b_m, b_n \} = 2 \de_{mn}$.
In terms of these operators, Eq.~\eqref{ham1} takes the form
\bea H &=& \frac{i}{2} ~\sum_{n=1}^{N-1} ~[~(\ga_0 \cos \phi - \De) ~b_{2n} 
b_{2n+1} \non \\
&& ~~~~~~~~~~~~-~ (\ga_0 \cos \phi + \De) ~b_{2n-1} b_{2n+2} \non \\
&& ~~~~~~~~~~~~+~ \ga_0 \sin \phi ~(b_{2n-1} b_{2n+1} + b_{2n} b_{2n+2}) ~] 
\non \\
&& + ~\frac{i}{2} ~\sum_{n=1}^N ~\mu ~b_{2n-1} b_{2n}, \label{ham2} \eea
up to a constant.
Note that the Hamiltonian is invariant under the parity transformation 
$\cal P$ corresponding to a reflection of the system about its mid-point, 
i.e., $b_{2n} \to (-1)^n b_{2N+1-2n}$ and $b_{2n+1} \to b_{2N-2n}$.

The Hamiltonian in Eq.~\eqref{ham2} has a time-reversal symmetry if $\ga$ 
is real, i.e., if $\phi = 0$ or $\pi$. The time-reversal transformation 
involves complex conjugating all numbers, including $i \to - i$, and
\beq b_{2n} ~\to~ -~ b_{2n} ~~~~{\rm and}~~~~ b_{2n+1} ~\to~ b_{2n+1}. 
\label{trs} \eeq
Note that $f_n$ and $f_n^\dg$ remain invariant under this transformation;
this implies that their Fourier transforms (defined below) transform as
$f_k \to f_{-k}$ and $f_k^\dg \to f_{-k}^\dg$.

The Hamiltonian in Eq.~\eqref{ham2} has particle-hole symmetry if 
$\mu = 0$. The particle-hole symmetry transforms $f_n \to (-1)^n 
f_n^\dg$, namely, 
\beq b_{2n-1} ~\to~ (-1)^n ~b_{2n-1} ~~~~{\rm and}~~~~ b_{2n} ~\to~ - ~(-1)^n
b_{2n}, \label{cph1} \eeq
and complex conjugates all numbers including $\ga \to \ga^*$.

It is convenient to define an operator
\bea F &=& \sum_{n=1}^N ~(2 f_n^\dg f_n - 1). \non \\
&=& -i ~\sum_{n=1}^N ~b_{2n-1} b_{2n}. \label{number} \eea
This is related to the total electron number, $\sum_n f_n^\dg f_n$, by some 
constants. We will see later how $F$ can be used to calculate the
average electron number of the end modes. 

The energy spectrum of Eq.~\eqref{ham2} in the bulk can be found by considering
a chain with periodic boundary conditions. We define the Fourier transform 
$f_k = \frac{1}{\sqrt N}~ \sum_{n=1}^N f_n e^{ikn}$, where the momentum $k$ 
goes from $-\pi$ to $\pi$ in steps of $2\pi/N$. Then Eq.~\eqref{ham1} can 
be written in momentum space as
\bea H &=& \sum_{0 \le k \le \pi} ~\left( \begin{array}{cc}
f_k^\dg & f_{-k} \end{array} \right) ~h_k ~\left( \begin{array}{c}
f_k \\
f_{-k}^\dg \end{array} \right), \label{ham3} \\
h_k &=& 2 \ga_0 \sin \phi \sin k ~I_2 ~+~ (2\ga_0 \cos \phi \cos k - \mu) ~
\tau^z \non \\
&& +~ 2 \De \sin k ~\tau^y, \label{hk1} \eea
where $I_2$ denotes the two-dimensional identity matrix and the $\tau^a$'s are 
Pauli matrices. The dispersion relation follows from 
Eqs.~(\ref{ham3}-\ref{hk1}) and is given by~\cite{gott2}
\bea E_{k\pm} &=& 2 \ga_0 \sin \phi \sin k \non \\
&& \pm ~\sqrt{(2\ga_0 \cos \phi \cos k - \mu)^2 ~+~ 4 \De^2 \sin^2 k}.
\label{disp} \eea

\begin{figure}[t] 
\begin{center}
\subfigure[]{\ig[width=3.4in]{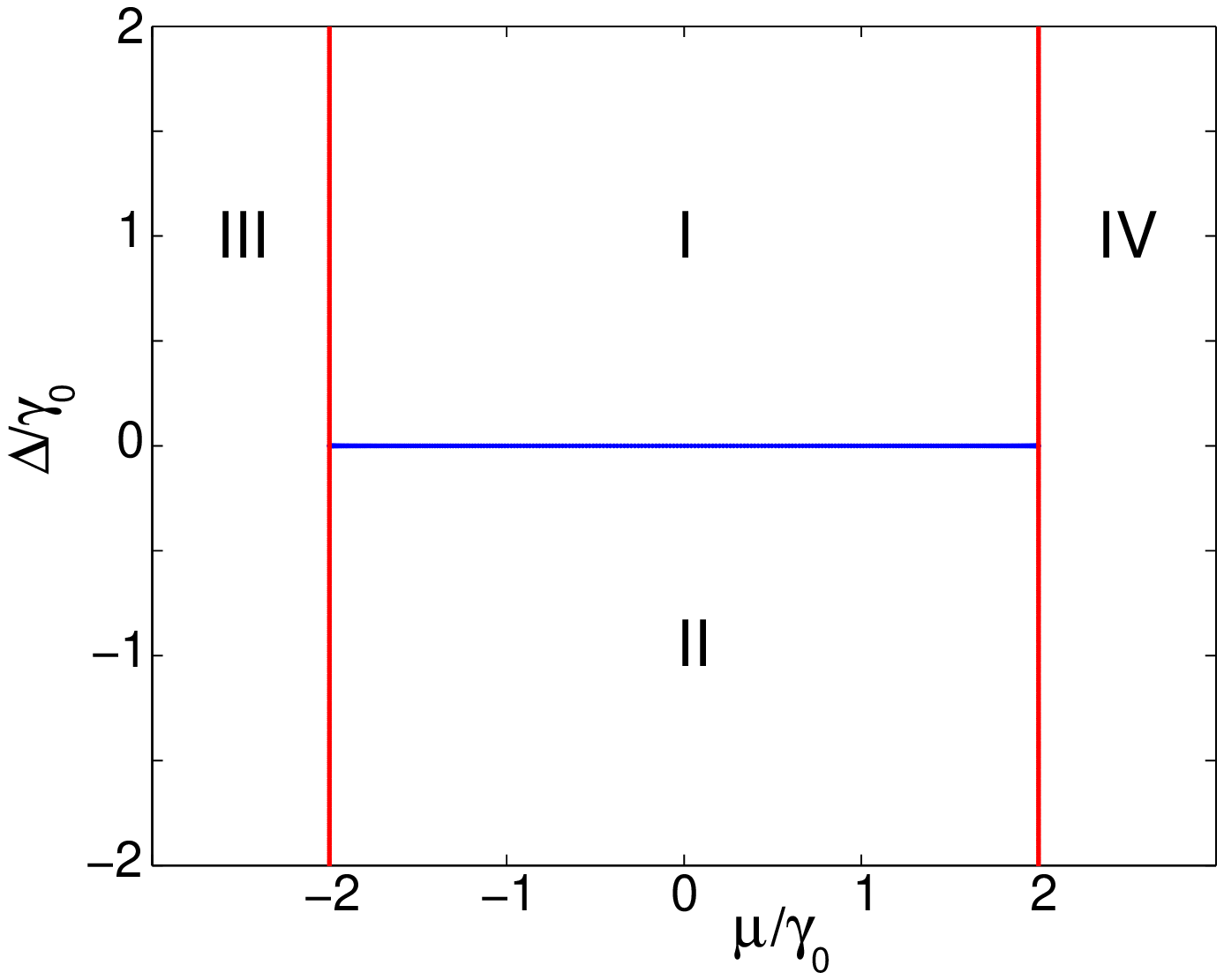}} \\
\subfigure[]{\ig[width=3.4in]{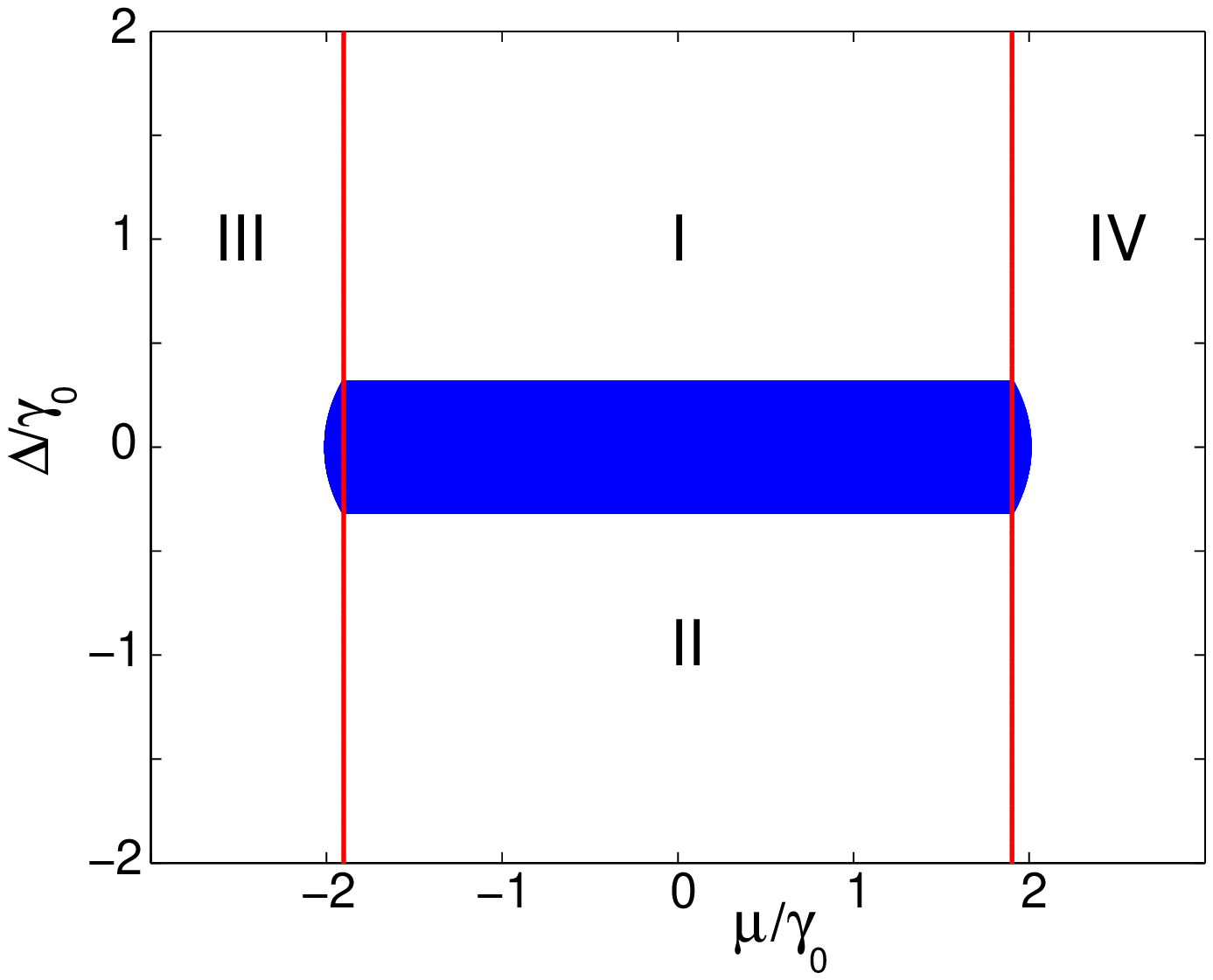}} 
\end{center}
\caption[]{Phase diagram of the model in Eq.~\eqref{ham2} as a function 
of $\mu/\ga_0$ and $\De/\ga_0$, for (a) $\phi = 0$ and (b) $\phi = \pi/10$. 
Phases I and II are topological with one mode at each end of a long chain, 
while III and IV are non-topological with no end modes; all these phases have
bulk spectra which are gapped. A blue shaded region appears in figure (b); in 
this region the bulk spectrum is gapless.} \label{fig:phase} \end{figure} 

Depending on the values of $\ga_0$, $\phi$, $\De$ and $\mu$, the system has 
four phases where $E_k$ is non-zero for all values of $k$, i.e., the bulk 
spectrum is gapped~\cite{gott2}. The phase 
diagram is shown in Fig.~\ref{fig:phase}. Phase I lies in the region 
$\De /\ga_0 > \sin \phi$ and $- 2 \cos \phi < \mu/\ga_0 < 2 \cos \phi$, while 
phase II lies in the region $\De /\ga_0 < - \sin \phi$ and $- 2 \cos \phi < 
\mu/\ga_0 < 2 \cos \phi$. In these two phases, a long and open chain has a 
zero energy mode at each end; hence these are called topological phases. 
Phases III and IV are non-topological phases in which there are no end modes.
In addition to these phases with a gapped bulk spectrum, we find a region in 
which the bulk spectrum is gapless if $\phi \ne 0$ or $\pi$. This is 
shown by the blue shaded 
region in Fig.~\ref{fig:phase} (b); this region consists of a rectangle which 
is bounded by the lines $\De/\ga_0 = \pm \sin \phi$ and $\mu /\ga_0 = \pm 2 
\cos \phi$ and is capped by two elliptical regions on the left and right sides.
There are no end modes in this region with a gapless bulk spectrum.

Next, we review the topological invariants which exist for a time-independent 
Hamiltonian of the form given in Eq.~\eqref{ham3}. This discussion will be 
useful for Sec.~\ref{sec:end} where we will study if similar topological 
invariants exist for a system in which the Hamiltonian varies periodically 
with time.

We consider a general form of $h_k$ in Eq.~\eqref{ham3} given by
\beq h_k ~=~ a_{0,k} ~I ~+~ a_{1,k} ~\tau^x ~+~ a_{2,k} ~\tau^y ~+~ a_{3,k} ~
\tau^z, \label{hk2} \eeq
where $k$ lies in the range $[0,\pi]$, and Hermiticity dictates that the 
$a_{i,k}$'s are all real functions of $k$. We assume that the bulk
spectrum is gapped for all values of $k$; since the energies are given by
$E_{k\pm} = a_{0,k} \pm \sqrt{a_{1,k}^2 + a_{2,k}^2 + a_{3,k}^2}$, all four
of the $a_{i,k}$'s cannot vanish simultaneously at any value of $k$.
Next, since $k$ is the same as 
$-k$ for $k=0$ and $\pi$, and the anticommutation relations imply that
$f_0 f_0 = 0$ and $f_0^\dg f_0 = - f_0 f_0^\dg$ plus a constant (and 
similarly for $k=\pi$), we can assume that $a_{0,k} = a_{1,k} = a_{2,k} = 0$ 
for $k=0, ~\pi$. Hence $a_{3,k}$ must be non-zero at $k= 0, ~\pi$, otherwise
the bulk spectrum would be gapless. It then turns out that the quantity $\nu 
= sgn (a_{3,0} a_{3,\pi})$ is a topological invariant; since it can only take 
values $\pm 1$, it is a $Z_2$-valued invariant. We find that a phase is 
topological (with an odd number of zero energy modes at each end of a long 
chain) if $\nu = -1$ and is non-topological (with either no end modes or 
an even number of zero energy end modes) if $\nu = +1$~\cite{gott2}.
The symmetry class of this general model which may not
have time-reversal symmetry is called class D.

If we impose time-reversal symmetry, we obtain additional constraints
on the $a_{i,k}$'s and a different topological invariant. If the Hamiltonian 
in Eq.~\eqref{hk2} has to be symmetric under the time-reversal transformation 
given in Eq.~\eqref{trs}, we must have $a_{0,k} = a_{1,k} = 0$ for all values 
of $k$. Hence $h_k$ only depends on two functions,
\beq h_k ~=~ a_{2,k} ~\tau^y ~+~ a_{3,k} ~\tau^z. \label{a23k} \eeq
Although Eq.~\eqref{ham3} defines $a_{2/3,k}$ only for $0 \le k \le \pi$, it 
is convenient to analytically continue these definitions to the entire range 
$-\pi \le k \le \pi$, with $a_{2,-k} = - a_{2,k}$ and $a_{3,-k} = a_{3,k}$.
Next we map 
$h_k$ to a vector $\vec V_k = a_{2,k} \hat y + a_{3,k} \hat z$ in the $y-z$ 
plane, and define the angle $\phi_k = \tan^{-1} (a_{3,k}/a_{2,k})$ made by 
the vector $\vec V_k $ with respect to the $\hat z$ axis. Following 
Ref.~\onlinecite{niu}, we now define a winding number 
as the integral over the Brillouin zone,
\beq W ~=~\int_{-\pi}^{\pi} ~\frac{dk}{2\pi} ~\frac{d\phi_k}{dk}. 
\label{wind} \eeq
This can take any integer value and is a therefore a $Z$-valued topological 
invariant. Note that this is well-defined since both $a_{2,k}$ and $a_{3,k}$ 
cannot simultaneously vanish at any value of $k$, otherwise the bulk spectrum 
would be gapless. A phase is topological if $W \ne 0$; such a phase will have 
$W$ zero energy modes at each end of a long chain. If $W = 0$, the phase is 
non-topological and does not have any end modes~\cite{gott2}. The 
symmetry class of this model with time-reversal symmetry is called class BDI.

We now consider the phase diagrams shown in Fig.~\ref{fig:phase}.
In Fig.~\ref{fig:phase} (a) we have time-reversal symmetry, and we find
that the winding number $W$ is equal to $-1$ in phase I, $+1$ in phase II,
and zero in phases III and IV. The invariant $\nu$ is equal to $-1$
in phases I and II and $+1$ in phases III and IV. In Fig.~\ref{fig:phase} (b) 
we do not have time-reversal symmetry; hence only the invariant $\nu$ is
defined. We find that $\nu$ is equal to $-1$ in phases I and II and $+1$ in 
phases III and IV.

\section{Floquet time evolution and end modes}
\label{sec:flo}

We will now begin our study of what happens when some parameter in the 
Hamiltonian is varied periodically in time. In this section we will
describe the numerical technique that we use to study the Floquet time
evolution and to find the end modes.

We consider a general Hamiltonian which is quadratic in terms of 
Majorana operators. For a system with $N$ sites, we have 
\beq H ~=~ \frac{i}{4} \sum_{m,n=1}^{2N} ~b_m M_{mn} b_n, \label{ham4} \eeq
where $M$ is a real antisymmetric matrix, so that $iM$ is Hermitian.
We allow $M$ to vary periodically with time so that $M(t+T)=M(t)$.
Equation \eqref{ham4} implies that the Heisenberg equations for the operators 
$b_n (t)$ are given by
\bea \frac{db_m(t)}{dt} &=& i ~[H(t),b_m (t)] \non \\
&=& \sum_{n=1}^{2N} ~ M_{mn} (t) ~b_n (t). \eea
If $b$ denotes the column vector $(b_1,b_2,\cdots,b_{2N})^T$ (the superscript 
$T$ denotes transpose), we can write the above equation as $db(t)/dt = M(t) 
b(t)$. The solution of this is given by
\bea b (t) &=& U(t,0) ~b(0), \non \\
{\rm where}~~ U(t,0) &=& {\cal T} e^{\int_0^t dt M(t)}, \label{ut} \eea
and $\cal T$ denotes the time-ordering symbol. The time evolution operator 
$U(t,0)$ can be numerically computed given the form of $M(t)$. Note that 
$U(T,0)$ is not only a unitary matrix, it is also real and orthogonal since 
$M(t)$ is real.

Since $M(t)$ varies with a time period $T$, we will call $U(T,0)$ the Floquet 
operator. The eigenvalues of $U(T,0)$ are given by phases $e^{i\ta_j}$ (where 
the $\ta_j$ lie in the range $[-\pi,\pi]$), and they come in complex conjugate 
pairs if $e^{i\ta_j} \ne 1$. This is because $U(T,0) \psi_j = e^{i\ta_j} 
\psi_j$ implies that $U(T,0) \psi_j^* = e^{-i\ta_j} \psi_j^*$ since $U(T,0)$ 
is real. For eigenvalues $e^{i\ta_j} = \pm 1$ (these eigenvalues may or may 
not have a degeneracy), the eigenvectors can be chosen to be real.

The Floquet operator $U(T,0)$ satisfies an additional property if the system 
has time-reversal symmetry. Time-reversal symmetry implies that we must have 
the matrix elements $M_{mn} = 0$ whenever $m-n$ is an even integer (see 
Eq.~\eqref{trs}), and $M(T-t)=M(t)$ (this imposes a restriction on the form of
the driving protocol). The first property combined with the antisymmetry of 
$M(t)$ implies that $\Sigma^z M^T (t) \Sigma^z = M (t)$, where $\Sigma^z$ is 
a diagonal matrix with 
\beq \Sigma^z_{2n-1,2n-1} ~=~ 1 ~~~~{\rm and}~~~~ \Sigma^z_{2n,2n} ~=~ -1. \eeq
(Note that $\Sigma^z$ is both unitary and Hermitian and satisfies $\Sigma^{z2} 
= I_{2N}$). We can then show that the Floquet operator $U(T,0)$ satisfies 
the relation 
\beq \Sigma^z U^T \Sigma^z ~=~ U. \label{trssym} \eeq
Since $U^T = U^{-1}$, Eq.~\eqref{trssym} implies that if $\psi_j$ is an 
eigenvector of $U$ with eigenvalue $e^{i\ta_j}$, $\Sigma^z \psi_j$ is an 
eigenvector of $U$ with eigenvalue $e^{-i\ta_j}$. Combining this with a 
statement made in the previous paragraph, we see that if the eigenvalue 
$e^{i\ta_j}$ is non-degenerate, the vectors $\psi_j^*$ and $\Sigma^z \psi_j$ 
must be identical up to a phase.

If the system has particle-hole symmetry, $U(T,0)$ satisfies the following
property. Following Eq.~\eqref{cph1}, we define a particle-hole transformation 
matrix $C$ which is diagonal with
\beq C_{2n-1,2n-1} ~=~ (-1)^n ~~~~{\rm and}~~~~ C_{2n,2n} ~=~ - ~(-1)^n.
\label{cph2} \eeq
($C$ is both unitary and Hermitian and satisfies $C^2 = I_{2N}$. 
Hence the eigenvalues of $C$ are $\pm 1$). Then particle-hole symmetry 
implies that 
\beq C U C ~=~ U. \label{cphsym} \eeq
This implies that eigenvectors of $U$ corresponding to non-degenerate
eigenvalues must necessarily be eigenvectors of $C$.

Following Eq.~\eqref{number} we define a matrix $\Sigma^y$ whose only 
non-zero elements are given by
\beq \Sigma^y_{2n-1,2n} ~=~ -i ~~~~{\rm and}~~~~ \Sigma^y_{2n,2n-1} ~=~ i. \eeq
Note that $\Sigma^y$ is Hermitian and has eigenvalues $\pm 1$. Hence,
in any state $\psi$, the expectation value $\psi^\dg \Sigma^y \psi$ must
lie between $-1$ and $1$. A state with $\psi^\dg \Sigma^y \psi = +1 ~(-1)$
is called a particle (hole) state respectively; this interpretation comes
from the fact that the operator in Eq.~\eqref{number} is related to the 
electron number.

We observe that the matrices $C$ and $\Sigma^y$ anticommute. This implies 
that if $\psi$ is an eigenvector of $C$, then the expectation value $\psi^\dg 
\Sigma^y \psi = 0$; this shows that such a state has equal probabilities of 
particles and holes.

In Sec.~\ref{sec:end}, we will consider two kinds of periodic driving of 
the hopping amplitude $\ga$. In each case, we will look for eigenvectors of 
$U(T,0)$ which are localized near the ends of the chain. Before discussing the
specific results in the next section, we will first describe our numerical 
method of finding the end modes and some of their general 
properties~\cite{thakur1}. 

The most convenient way of finding eigenvectors of $U(T,0)$ which 
are localized at the ends is to look at the inverse participation ratio 
(IPR). We assume that the eigenvectors, denoted as $\psi_j$, are 
normalized so that $\sum_{m=1}^{2N} |\psi_j (m)|^2 = 1$ for each value of
$j$; here $m=1,2,\cdots,2N$ labels the components of the eigenvector. 
The IPR of an eigenvector is then defined as $I_j = \sum_{m=1}^{2N} |\psi_j 
(m)|^4$. If $\psi_j$ is extended equally over all sites so that 
$|\psi_j (m)|^2 = 1/(2N)$ for each $m$, then $I_j = 1/(2N)$ and
this will approach zero as $N \to \infty$. But if $\psi_j$ is localized over
a distance $\xi$ (which is of the order of the decay length of the eigenvector
and remains constant as $N \to \infty$), then we will have $|\psi_j (m)|^2 
\sim 1/\xi$ in a region of length $\xi$ and $\sim 0$ elsewhere; then we obtain 
$I_j \sim 1/\xi$ which will remain finite as $N \to \infty$. Hence, if $N$ is 
sufficiently large, a plot of $I_j$ versus $j$ will allow us to distinguish 
between states which are localized and extended states. Once we find a 
state $j$ for which $I_j$ is significantly larger than $1/(2N)$, we look at 
a plot of the probabilities $|\psi_j (m)|^2$ versus $m$ to see whether it is 
indeed an end state. Finally, we check if the form of $|\psi_j (m)|^2$ and 
the value of its IPR remain unchanged if $N$ is increased. We find that
the IPR of an end mode saturates to a constant value once $N$ becomes larger 
than about twice its decay length $\xi$. We will not show a plot of 
the IPR versus $N$ here since this is rather simple.

In the periodic driving protocols discussed in Sec.~\ref{sec:end}, we 
find that, for certain ranges of the parameter values, $U(T,0)$ has one or 
more pairs of eigenvectors with substantial values of the IPR. We find that
each such pair corresponds to modes localized at the two ends of the system.
Further, the eigenvalues of such a pair become degenerate in the limit
that the system size $N$ is much larger than the decay length $\xi$ of
the end modes. The existence of such pairs of eigenstates follows
from the parity symmetry of the Hamiltonian discussed after Eq.~\eqref{ham2}
which leads to a similar symmetry of $U(T,0)$. Namely, if $\psi_1$ is 
an eigenstate of $U(T,0)$ which is localized near one end of the system, 
the parity transformation gives an eigenstate $\psi_2 = {\cal P} \psi_1$ 
which is localized near the other end. The eigenvalues are degenerate
in the limit $N \gg \xi$; if $N \lesssim \xi$, there is tunneling 
between the two end modes and this breaks the degeneracy.

\section{Periodic driving of hopping amplitude}
\label{sec:end}

In this section we will study in detail two cases which correspond 
respectively to the magnitude and the phase of the hopping amplitude
varying sinusoidally with a time period $T$. Namely, we will consider \\
\noi (i) $\ga (t) ~=~ \ga_0 ~[1 ~+~ a \cos (\om t)]$, and \\
\noi (ii) $\ga (t) = ~\ga_0 ~e^{i a \cos (\om t)}$, \\ 
where $a$ is real and $\om = 2\pi/T$. Physically we may think of case (i) as 
arising from a periodic application of pressure on the system. This would make
the lattice spacing and therefore the strength of the hopping vary 
with time. Case (ii) can arise due to the application of electromagnetic 
radiation on the system. This gives rise to an electric field and therefore 
a vector potential which varies sinusoidally in time. The vector potential
can be put into the phase of the hopping by the Peierls prescription.

Using Eqs.~\eqref{ham2} and \eqref{ham4}, we find that the matrix elements 
of $M(t)$ are given by 
\bea M_{2n,2n+1} &=& - ~M_{2n+1,2n} ~=~ \ga_R (t) ~-~ \De, \non \\
M_{2n-1,2n+2} &=& - ~M_{2n+2,2n-1} ~=~ -~ \ga_R (t) ~-~ \De, \non \\
M_{2n-1,2n+1} &=& - ~M_{2n+1,2n-1} ~=~ \ga_I (t), \non \\
M_{2n,2n+2} &=& - ~M_{2n+2,2n} ~=~ \ga_I (t), \non \\
M_{2n-1,2n} &=& - ~M_{2n,2n-1} ~=~ \mu, \label{mt2} \eea
where $n$ runs over appropriate ranges of values in the different
equations, and $\ga_R (t)$ and $\ga_I
(t)$ denote the real and imaginary parts of $\ga (t)$. Namely, $\ga_R = \ga_0 
[1 + a \cos (\om t)]$ and $\ga_I = 0$ in case (i), while $\ga_R = \ga_0 \cos 
(a \cos (\om t))$ and $\ga_I = \ga_0 \sin (a \cos (\om t))$ in case (ii).
We then numerically calculate the Floquet operator 
\beq U(T,0) ~=~ {\cal T} e^{\int_0^T dt M(t)}, \eeq
find all its eigenstates and eigenvalues, and use the IPR to identify the
end modes as explained in Sec.~\ref{sec:flo}.

\begin{figure}[htb] 
\subfigure[]{\includegraphics[width=8.6cm]{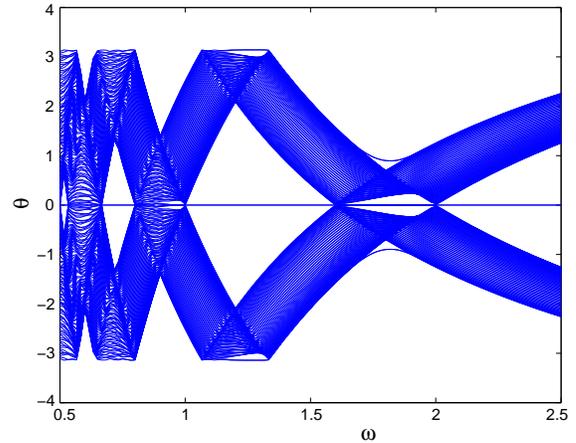}}
\subfigure[]{\includegraphics[width=8.6cm]{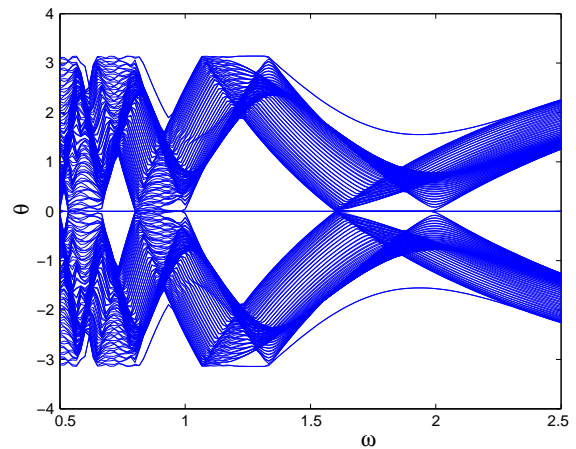}} \\
\caption[]{Values of $\ta$ (lying in the range $[-\pi,\pi]$) versus $\om$,
for a 200-site open chain with $\ga_0 =1$, $\De = 0.8$, $\mu = 0$, and
$\ga (t) = \ga_0 [1 + a \cos (\om t)]$. In (a) $a=0.5$, and in (b) $a=1$.
The isolated lines correspond to end modes: the ones with $\ta = 0$ and $\pm 
\pi$ are conventional end modes, while the ones with $\ta \ne 0$ or $\pm \pi$ 
are anomalous end modes.} \label{fig:quasi_real} \end{figure}

\begin{figure}[htb] 
\subfigure[]{\includegraphics[width=8.6cm]{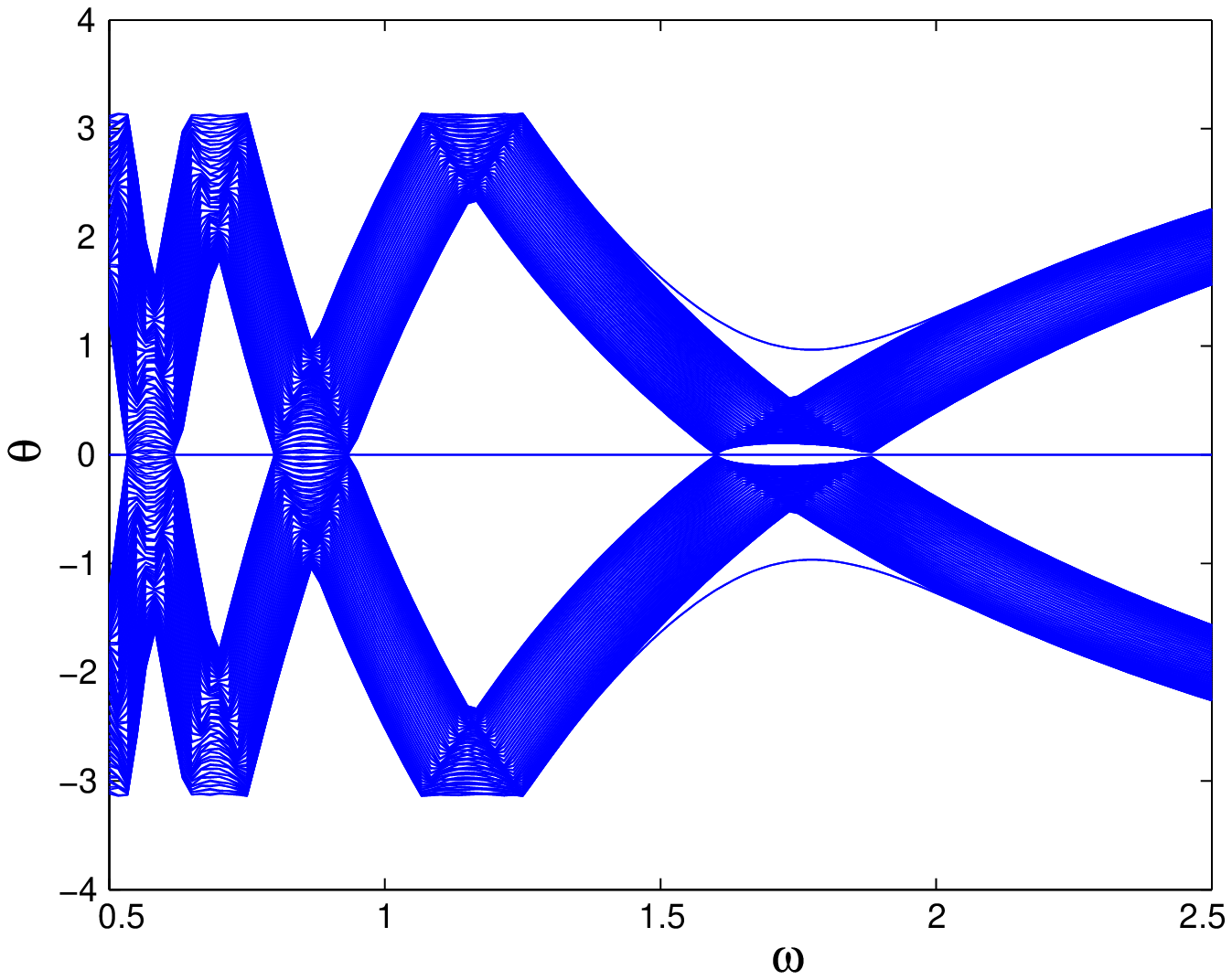}}
\subfigure[]{\includegraphics[width=8.6cm]{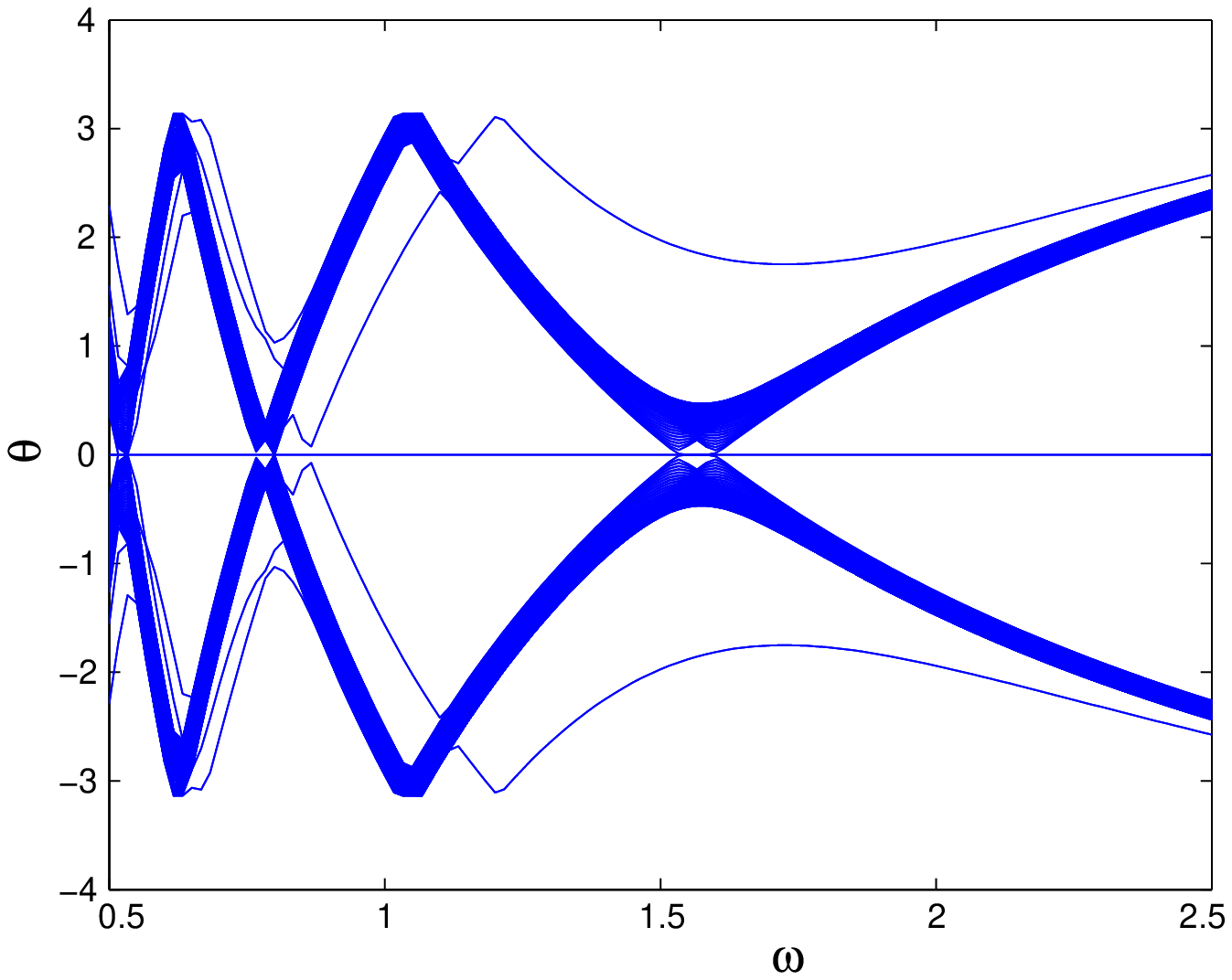}} \\
\caption[]{Values of $\ta$ (lying in the range $[-\pi,\pi]$) versus $\om$,
for a 200-site open chain with $\ga_0 =1$, $\De = 0.8$, $\mu = 0$, and
$\ga (t) = \ga_0 e^{i a \cos (\om t)}$. In (a) $a=0.5$, and in (b) $a=1$.
The isolated lines correspond to end modes: the ones with $\ta = 0$ and $\pm 
\pi$ are conventional end modes, while the ones with $\ta \ne 0$ or $\pm \pi$ 
are anomalous end modes.} 
\label{fig:quasi_imag} \end{figure}

We will now present our numerical results. For most of our studies, we 
consider a 200-site open chain (hence with a 400-dimensional Hamiltonian) with
$\ga_0 =1$, $\De = 0.8$, and $\mu = 0$. We begin with a quick view of 
the Floquet eigenvalues $e^{i\ta_j}$ of all the modes of the system.
In Fig.~\ref{fig:quasi_real} we present the values of $\ta_j$ as a function 
of the driving frequency $\om$ for case (i) with $\ga (t) ~=~ \ga_0 ~[1 ~+~ a 
\cos (\om t)]$ for $a = 0.5$ and 1. Figure \ref{fig:quasi_imag} shows the 
values of $\ta_j$ versus $\om$ for case (ii) with $\ga (t) = ~\ga_0 ~e^{i 
a \cos (\om t)}$ for $a = 0.5$ and 1. [The Floquet eigenvalues are sometimes
written as $e^{i\ta_j} = e^{-i \ep_j T}$, where $\ep_j$ are called the
quasienergies; these lie in the range $[-\pi/T,\pi/T]$. However, we will 
generally work with the variable $\ta_j$ rather than $\ep_j$ due to the 
simplifying feature that the range of $\ta_j$ does not depend on 
$T$]. In Figs.~\ref{fig:quasi_real} and \ref{fig:quasi_imag}, we see
some continuous bands and some isolated lines which are separated from the
bands for certain ranges of $\om$. The bands turn out to consist 
of bulk modes whose wave functions are spread throughout the system, while
the isolated lines correspond to end modes whose wave functions are localized
near the two ends of the system. For some particular values of the system 
parameters, we have confirmed that the modes with isolated Floquet eigenvalues
are end modes by looking at the IPRs of all the eigenvectors of the
Floquet operator, picking out the ones whose IPRs are larger by a factor
of 2 or more than the remaining ones, and looking at the wave functions
of these modes to check that they are localized at the ends. We find 
numerically that as the separation between an end mode and the bulk band 
decreases, the decay length of the mode from the end of the chain increases 
and hence its IPR decreases.


In Figs.~\ref{fig:quasi_real} (a) and \ref{fig:quasi_imag} (a), we 
see that the number of end modes show several 
changes in the interval $1.6 \le \om \le 2$ when the driving amplitude 
$a$ is small. We can qualitatively understand this as follows.
In the absence of driving (when $a=0$), the parameters $\ga_0 = 1, ~\De = 0.8$
and $\mu = 0$ place the system in the topological phase I in 
Fig.~\ref{fig:phase} (a). An open chain then has a mode at each end with
zero energy, while the bulk bands lie in the ranges $[-2,-1.6]$ and $[1.6,2]$
according to Eq.~\eqref{disp}. Hence, periodic driving with a frequency
$\om$ can, to first order in $a$, produce transitions between an end mode
and the bulk states if $\om$ lies in the range $[1.6,2]$. This explains why 
so many changes in the end modes occur in this range of frequencies.

\begin{figure}[htb] 
\subfigure[]{\includegraphics[width=8.6cm]{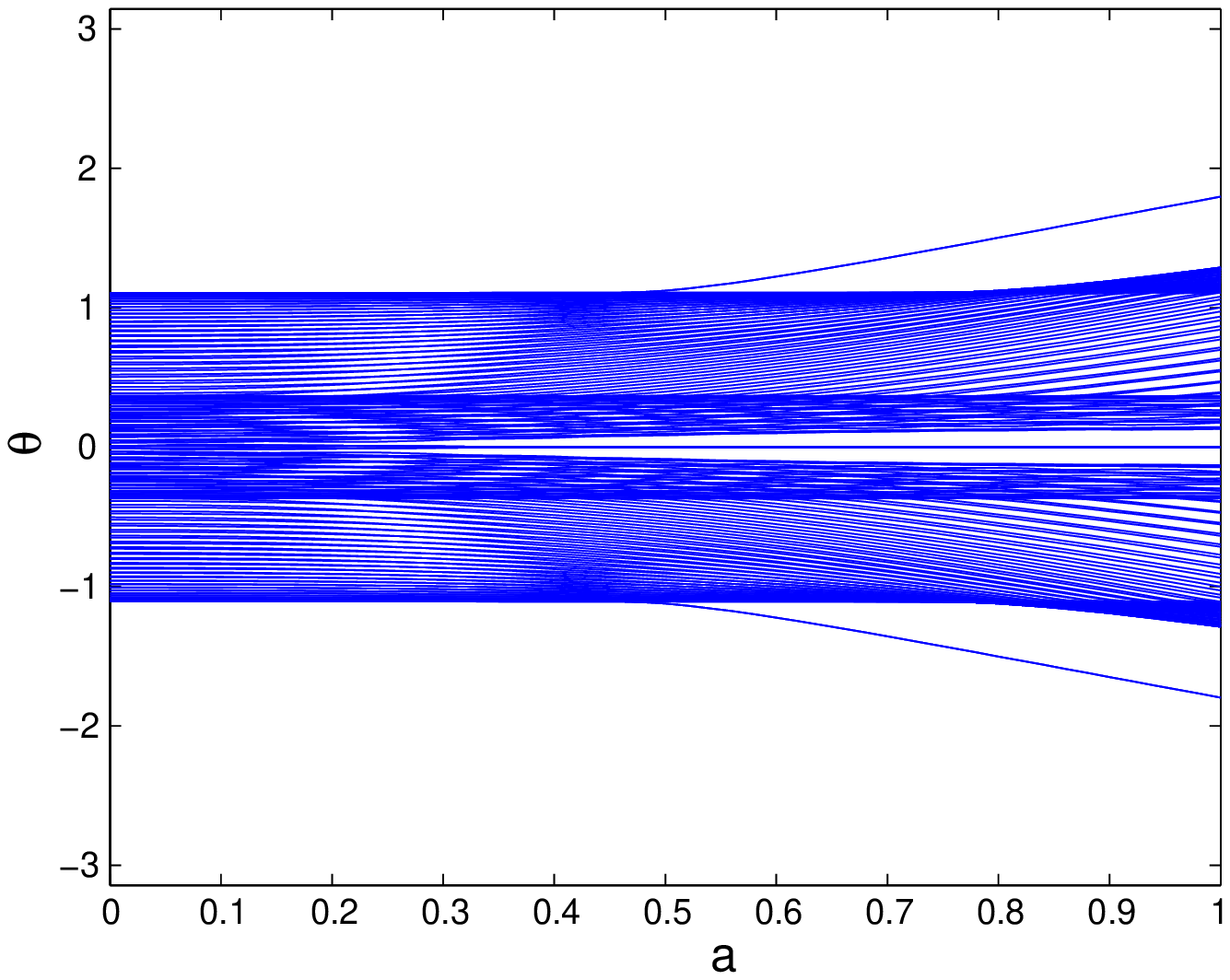}}
\subfigure[]{\includegraphics[width=8.6cm]{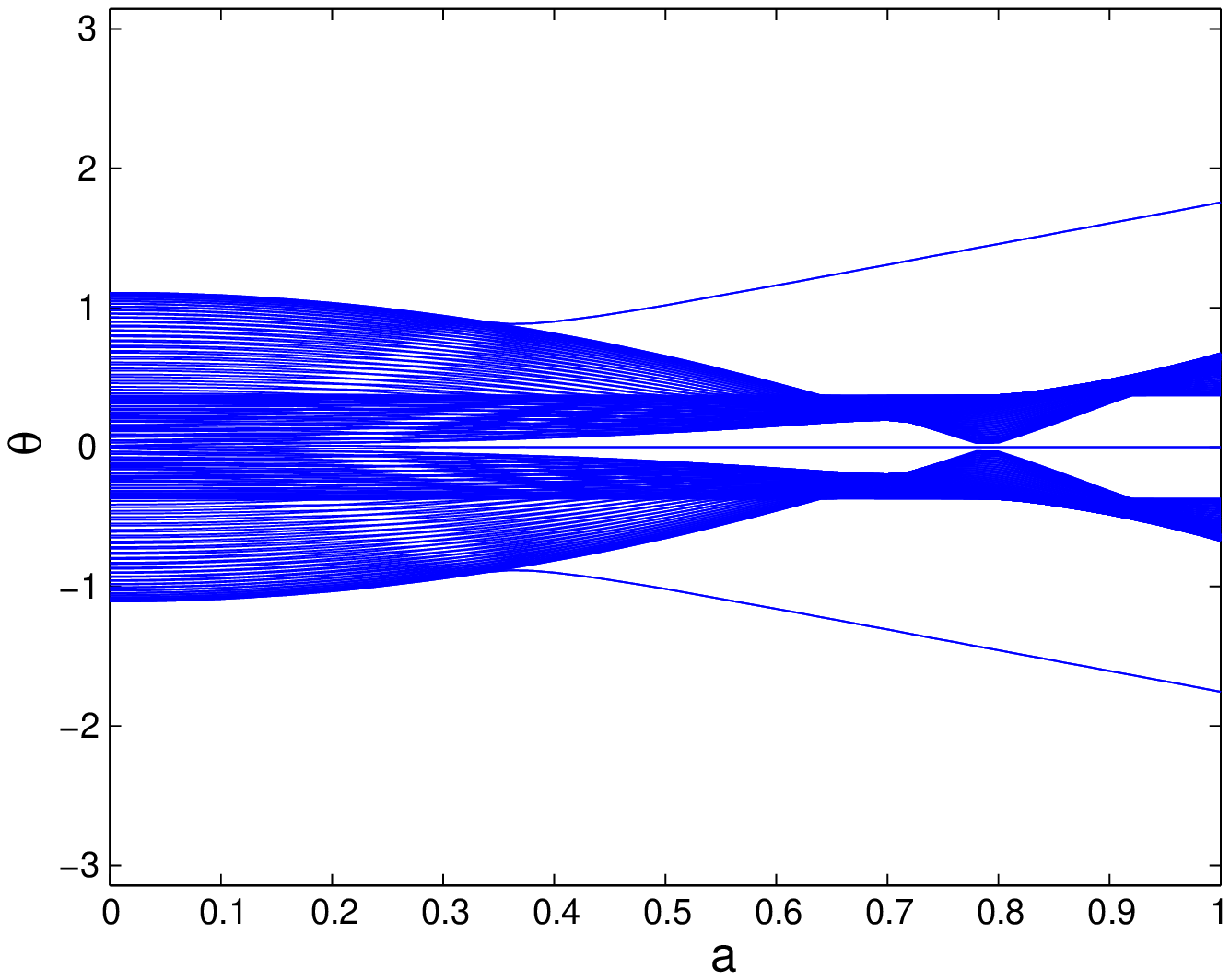}} \\
\caption[]{Values of $\ta$ (lying in the range $[-\pi,\pi]$) versus $a$, for 
a 200-site open chain with $\ga_0 =1$, $\De = 0.8$, $\mu = 0$, and $\om =1.7$.
In (a) $\ga (t) = \ga_0 [1 + a \cos (\om t)]$, and in (b) $\ga (t) = \ga_0 
e^{i a \cos (\om t)}$. The isolated lines correspond to end modes: the ones 
with $\ta = 0$ and $\pm \pi$ are conventional end modes, while the ones with 
$\ta \ne 0$ or $\pm \pi$ are anomalous end modes.} 
\label{fig:quasi_amp} \end{figure}

In Fig.~\ref{fig:quasi_amp}, we show $\ta_j$ as a function of the driving
amplitude $a$ for a 200-site open chain with $\ga_0 =1$, $\De = 0.8$, 
$\mu = 0$, and $\om =1.7$. We again see some continuous bands and some 
isolated lines corresponding to end modes. It is clear that anomalous end 
modes with $\ta \ne 0$ or $\pm \pi$ appear only when $a$ is sufficiently far
from zero.

\begin{figure}[htb] \ig[width=3.4in]{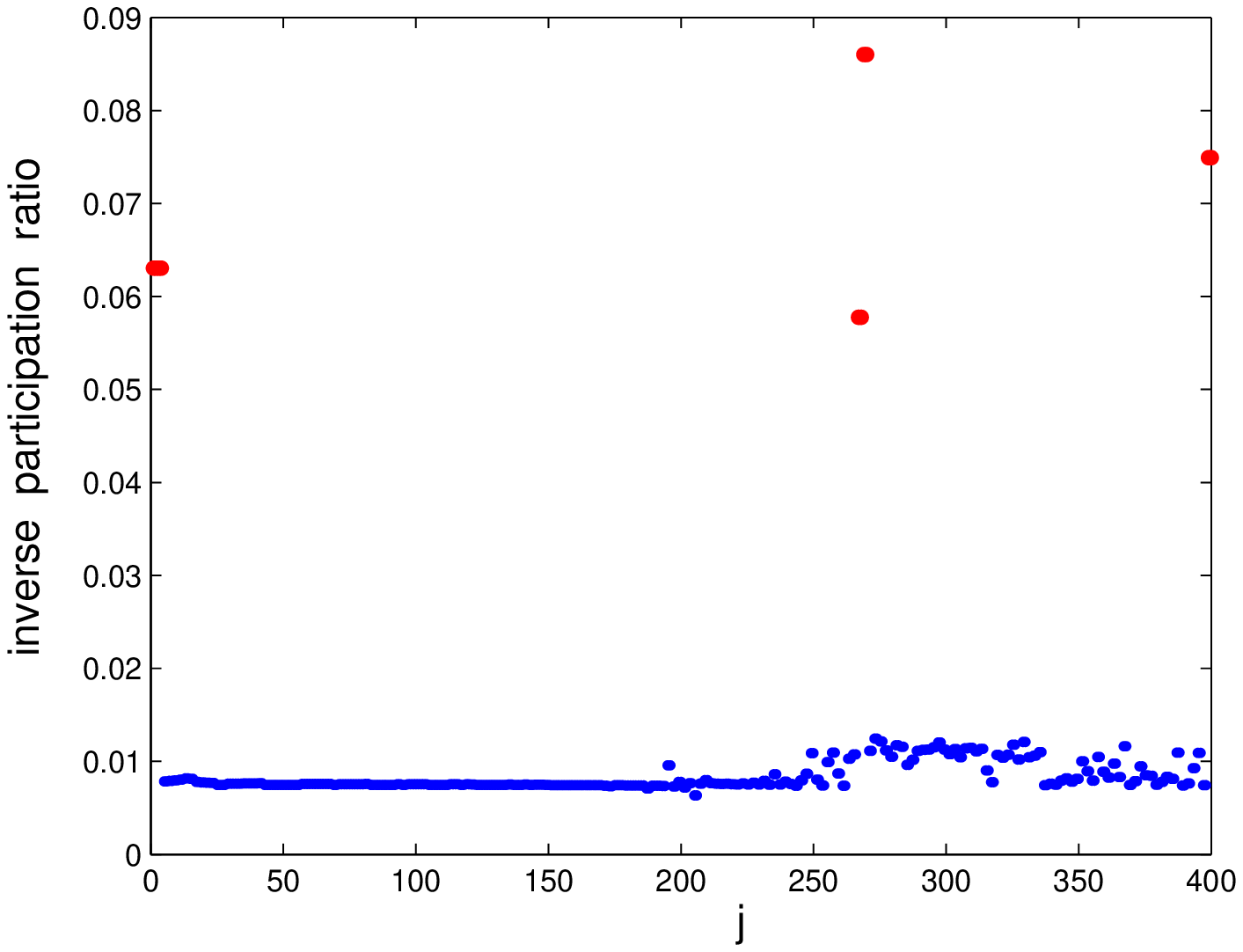} 
\caption[]{IPRs of different eigenvectors of the Floquet operator for a 
200-site open chain with $\ga_0 =1$, $\De = 0.8$, $\mu = 0$, $a=0.5$ and 
$\om =1.7$. There are ten modes with large IPRs shown as large red dots; their
IPR values (and their degeneracies in parentheses) are $0.0860$ (two), $0.0749$
(two), $0.0630$ (four), and $0.0578$ (two).} \label{fig:ipr} \end{figure}

\begin{figure}[htb] \ig[width=3.4in]{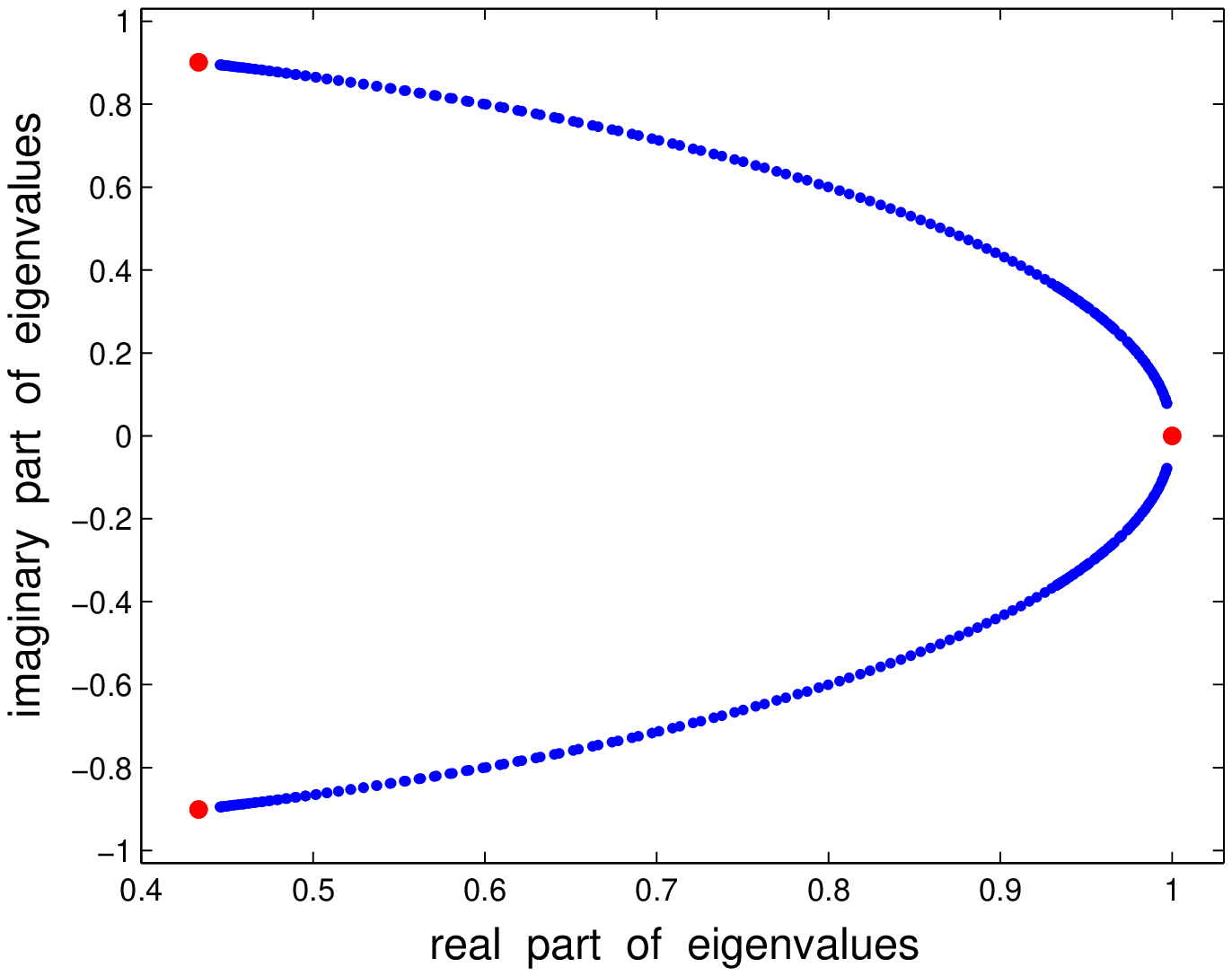}
\caption[]{Real and imaginary parts of different eigenvalues of the Floquet 
operator for a system with the same parameters as in Fig.~\ref{fig:ipr}. Ten
isolated eigenvalues are visible (large red dots) corresponding to the 
eigenvectors with large IPRs in Fig.~\ref{fig:ipr}; the eigenvalues (and 
their degeneracies) are given by 1 (six) and $0.4333 + 0.9012 i$ (two), and 
$0.4333 - 0.9012 i$ (two). (There are no eigenvalues with real part less 
than $0.4333$).} \label{fig:floeig} \end{figure}

We now examine one particular case in detail to understand various aspects
of the problem. We consider case (i), $\ga(t) = \ga_0 [1 + a \cos (\om t)]$,
and we take the driving parameters to be $\om =1.7$ and $a=0.5$. In 
Fig.~\ref{fig:ipr} we show the IPRs of the 400 eigenvectors of the Floquet
operator $U(T,0)$ in increasing order. We find that there are ten modes whose 
IPRs are much larger than all the others. The values of these ten IPRs and 
their degeneracies are given in the caption of that figure. 
Figure~\ref{fig:floeig} shows the real and imaginary parts of all the 
eigenvalues of the Floquet operator; all these eigenvalues are of the form 
$e^{i\ta_j}$ and lie on the unit circle. We see that all the eigenvalues 
(except for ten) form two continuous bands, one with imaginary part positive 
and the other with imaginary part negative; the upper band of eigenvalues goes 
from $0.4458 + 0.8951 i$ to $0.9971 + 0.0761 i$ (namely, $\ta$ goes from
$0.0762$ to $1.1087$) while the lower band goes from $0.4458 - 0.8951 i$ to 
$0.9971 - 0.0761 i$ ($\ta$ goes from $-1.1087$ to $-0.0762$). The remaining 
ten eigenvalues lie outside these bands; six of them have eigenvalue 1 
(i.e., $\ta = 0$) while there are two each with eigenvalues $0.4333 \pm 
0.9012 i$ ($\ta = \pm 1.1226$). These ten states correspond precisely to the 
eigenvectors with the largest IPRs.

Looking at these ten eigenvectors, we find that they are all localized near 
the two ends of the chain. The four anomalous eigenvectors with eigenvalues
equal to $0.4333 \pm 0.9012 i$ have non-zero components $\psi_j (m)$ for
both even and odd values of $m$ at both ends (we recall that $m$ goes
from 1 to 400). However, the six eigenvectors with eigenvalues equal to 1 have
non-zero components only for even values of $m$ near the left end of the chain 
(i.e., near $m=1$) and only for odd values of $m$ near the right end of the 
chain (near $m=400$). End modes with Floquet eigenvalues equal to $\pm 1$
are sometimes called Floquet Majorana modes~\cite{jiang,tong,thakur1,kundu}; 
they are the time-dependent analogs of the Majorana end modes which appear in 
time-independent systems with zero energy~\cite{gott2}.

We find that all the ten end modes are eigenvectors of the matrix $C$
defined in Eq.~\eqref{cph2}. Hence the expectation value of $\Sigma^y$ is
zero in all these modes, and each mode therefore has equal probabilities of 
particles and holes. (This is exactly the property that zero energy Majorana
modes at the ends of an open chain have in a time-independent 
system~\cite{sengupta}).

\begin{figure}[htb] \ig[width=3.4in]{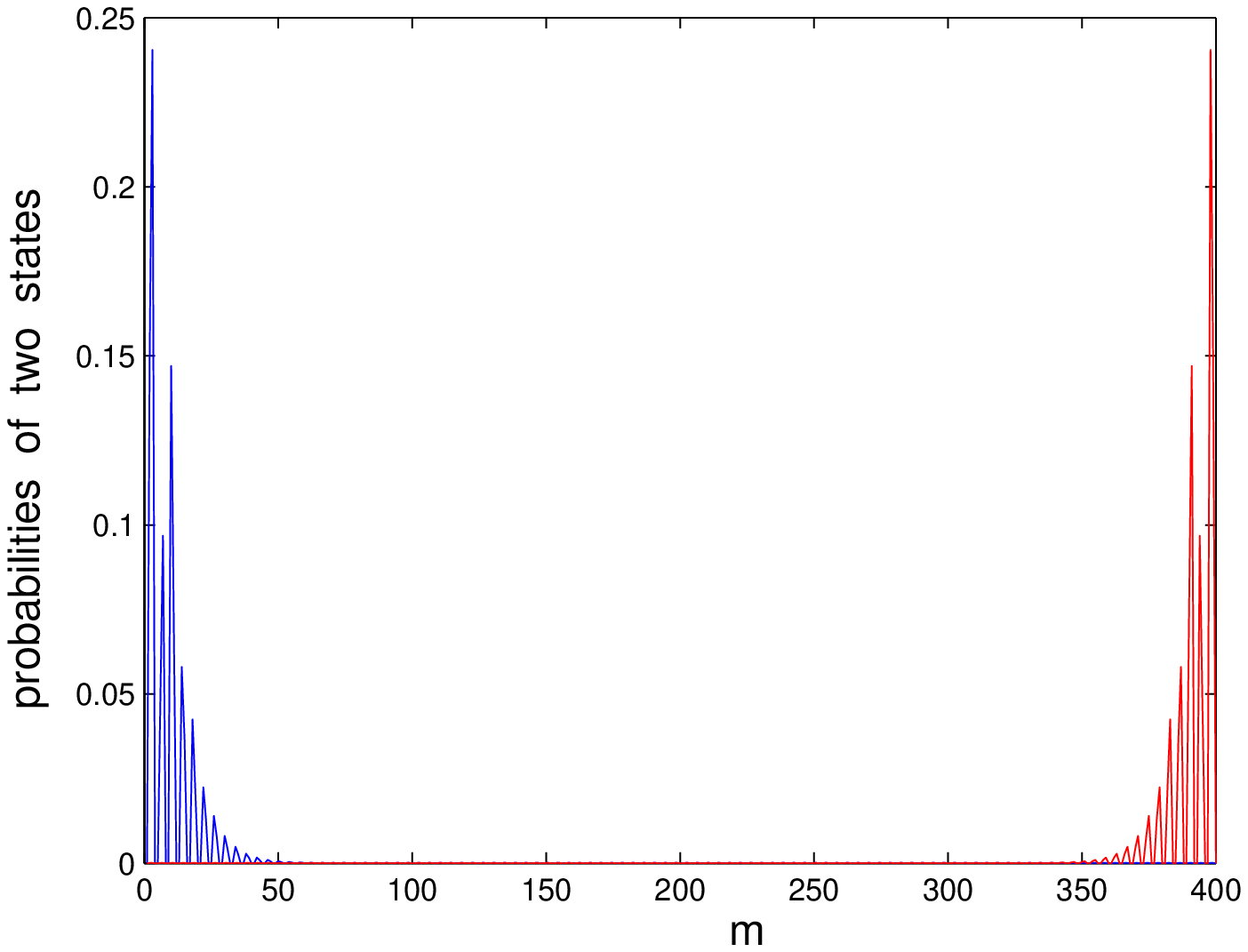}
\caption[]{Probabilities of two eigenstates of the Floquet operator which are 
localized at the two ends of the system; the parameter values are the same 
as in Fig.~\ref{fig:ipr}. Both states have Floquet eigenvalue equal to $0.4333
+ 0.9012 i$.} \label{fig:prob1} \end{figure}

\begin{figure}[htb] \ig[width=3.4in]{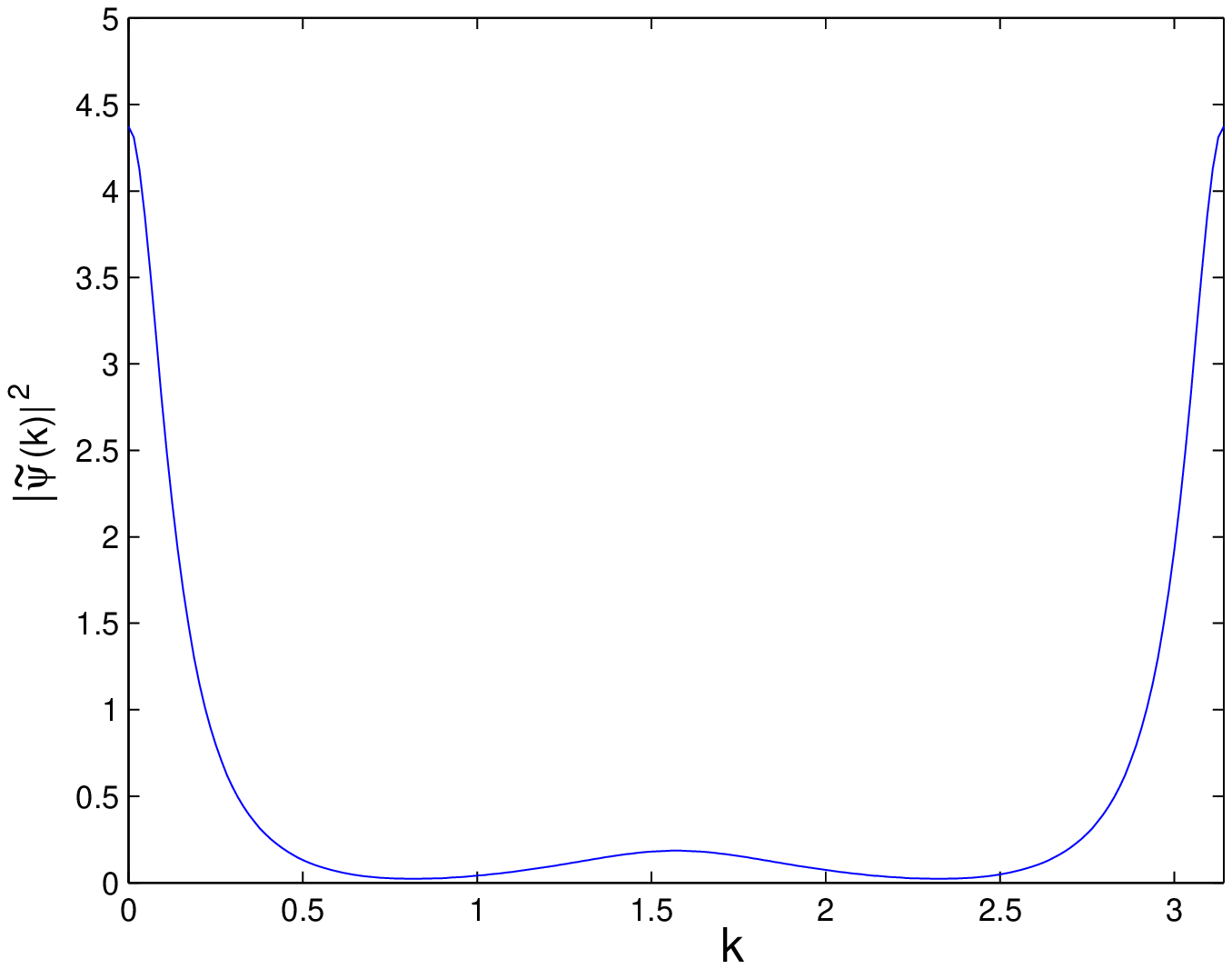}
\caption[]{$|\tilde{\psi_j} (k)|^2$ versus $k$ for the state localized at the 
left end of the chain with Floquet eigenvalue equal to $0.4333 + 0.9012 i$;
the parameter values are the same as in Fig.~\ref{fig:ipr}.} 
\label{fig:fourier1} \end{figure}

In Fig.~\ref{fig:prob1}, we show the probabilities $|\psi_j (m)|^2$ versus $m$
for two eigenvectors localized at the ends, both of which have Floquet 
eigenvalue equal to $0.4333 + 0.9012 i$. Next, we look at the Fourier 
transforms $\tilde{\psi_j} (k)$ of these wave functions. The 
Fourier transform can be defined for either odd or even numbered
sites, namely, as $\tilde{\psi_j} (k) = \sum_{n=1}^N \psi_j (2n-1) e^{ikn}$
or $\sum_{n=1}^N \psi_j (2n) e^{ikn}$. For the state localized at the left 
end of the chain, we show $|\tilde{\psi_j} (k)|^2$ (for the even numbered 
sites) versus $k$ for $0 \le k \le \pi$ in Fig.~\ref{fig:fourier1}. [The 
figure looks identical for $-\pi \le k \le 0$ since $|\tilde{\psi_j} (-k)|^2 
= |\tilde{\psi_j} (k)|^2$. Further, the figure looks similar, though not 
identical, for the Fourier transform of the odd numbered sites]. 
We find that the Fourier transform is peaked at $k=0$ and $\pi$. Similar 
results are found for the state localized at the right end.

\begin{figure}[htb] \ig[width=3.4in]{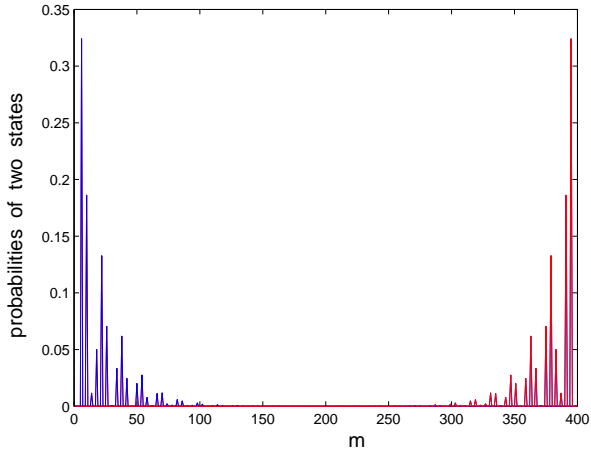}
\caption[]{Probabilities of two eigenstates of the Floquet operator which are 
localized at the two ends of the system; the parameter values are the same 
as in Fig.~\ref{fig:ipr}. Both states have Floquet eigenvalue equal to 1.}
\label{fig:prob2} \end{figure}

\begin{figure}[htb] \ig[width=3.4in]{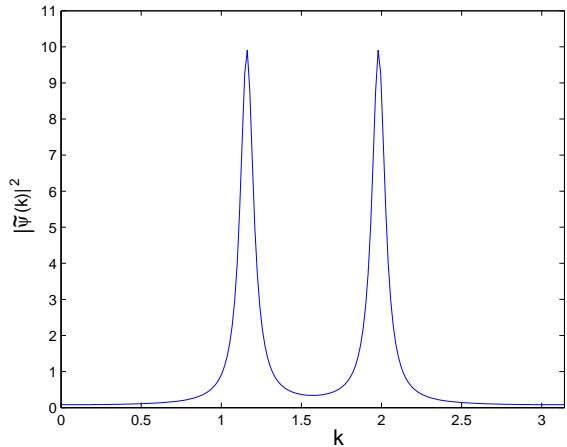}
\caption[]{$|\tilde{\psi_j} (k)|^2$ versus $k$ for the state localized at the 
left end of the chain with Floquet eigenvalue equal to 1; the parameter values 
are the same as in Fig.~\ref{fig:ipr}.} \label{fig:fourier2} \end{figure}

Similarly, in Fig.~\ref{fig:prob2}, we show the probabilities $|\psi_j (m)|^2$
versus $m$ for two eigenvectors localized at the ends, both of which have 
Floquet eigenvalue equal to 1. For the state localized at the left end, the 
wave function $\psi_j (m)$ is non-zero only for even values of $m$; we define 
its Fourier transform as $\tilde{\psi_j} (k) = \sum_{n=1}^N \psi_j (2n) 
e^{-ikn}$. Figure \ref{fig:fourier2} shows $|\tilde{\psi_j} (k)|^2$ versus $k$ 
for $0 \le k \le \pi$. We find that the Fourier transform is peaked at two 
values given by $k = 1.162$ and $1.979$; we note that these two values add up 
to $\pi$. Similar results are found for the state localized at the right end.

We have checked that the results presented in Figs. \ref{fig:ipr} to 
\ref{fig:fourier2} for the end modes (namely, 
the existence of ten such modes, their Floquet eigenvalues, wave functions,
and the locations of the peaks of their Fourier transforms) remain 
unchanged if the system size is increased from 200 to, say, 300.

\begin{figure}[htb] \ig[width=3.4in]{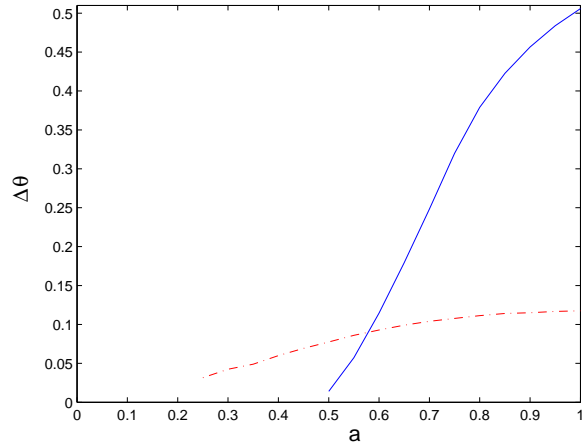}
\caption[]{Gap between the end of a continuous band of Floquet eigenvalues 
and the Floquet eigenvalue of an end mode versus $a$ for a 200-site open chain 
with $\ga_0 = 1, ~\De = 0.8, ~\mu = 0$ and $\om = 1.7$. The solid blue line
shows the gap between the end of a continuous band and an anomalous end mode 
(which appears for $a \gtrsim 0.5$), and the red dash-dotted line shows the gap
between the end of a continuous band and an end mode with Floquet eigenvalue 1 
(which appears for $a \gtrsim 0.25$).} \label{fig:gap} \end{figure}

In Fig.~\ref{fig:floeig} we see some gaps between the ends of the 
continuous bands of Floquet eigenvalues and the isolated Floquet eigenvalues 
of the end modes. We have studied how these gaps vary with the driving 
amplitude $a$. Since the Floquet eigenvalues are of the 
form $e^{i\ta}$, we define a gap as $\De \ta = |\ta_1 - \ta_2|$,
where $\ta_1$ is the eigenvalue at the end of a continuous band
and $\ta_2$ is the eigenvalue for an end mode. Figure \ref{fig:gap} shows 
the gap $\De \ta$ between the end of a continuous band and an
anomalous end mode (solid blue line) and an end mode with Floquet 
eigenvalue 1 (red dash-dotted line). This figure implies that a 
significantly larger driving amplitude is required to produce the anomalous
end modes compared to the end modes with Floquet eigenvalue 1.

We have studied what happens to the end modes if the chemical potential 
$\mu$ is not equal to zero. (This breaks the particle-hole
symmetry discussed in Eqs.~\eqref{cph1} and \eqref{cph2}). We have studied
a 200-site open chain with $\ga (t) = \ga_0 [1 + a \cos (\om t)]$, where
$\ga_0 = 1$, $\De = 0.8$, $\om = 1.7$, and $a = 0.5$. Upon varying $\mu$, we 
discover that while the three modes at each end with Floquet eigenvalue equal 
to $+1$ survive up to quite large values of $\mu$, the anomalous end modes 
with Floquet eigenvalues far from $\pm 1$ disappear as $\mu$ 
goes away from zero. To be precise, we find that the anomalous modes no 
longer appear when $|\mu|$ is larger than about $0.005$. We can understand 
this small number as follows. From the earlier discussion of the Floquet 
eigenvalues shown in Fig.~\ref{fig:floeig}, we know that the gap between the 
Floquet eigenvalues of the anomalous end modes and the ends of the bulk bands 
is given by $\De \ta = |\cos^{-1} (0.4333) - \cos^{-1} (0.4458)| =
0.0139$. This corresponds to a quasienergy gap equal to $\De \ep = \De \ta/T
= 0.0139 ~(\om/2\pi) = 0.0038$. We therefore expect that a perturbation like
$\mu$ will close the gap and the anomalous end modes will disappear if $\mu$ 
is of the order of $\De \ep$. We see that this gives the correct order of 
magnitude of the value of $\mu \simeq 0.005$ beyond which there are no 
anomalous end modes. To conclude, the existence of an anomalous end mode is 
sensitively dependent on $\mu$ being close to zero, with the critical value 
of $\mu$ being of the order of the quasienergy gap between the anomalous 
mode and the nearest end of a bulk band.

Before ending this section, we briefly comment about what happens for 
periodic driving in case (ii), with $\ga (t) = \ga_0 e^{i a \cos (\om t)}$.
We have studied in detail a 200-site open chain with $\ga_0 = 1, ~\De = 0.8, ~
\mu = 0, ~\om = 1.7$ and $a=0.4$. We then find six end modes, three at each
end. Of the three modes, two have eigenvalues $0.6216 \pm 0.7833 i$ 
(namely, anomalous modes with $\ta = \pm 0.9000$) and one has eigenvalue 1
(i.e., $\ta = 0$). All these eigenvalues lie outside the 
range of the Floquet eigenvalues of the bulk modes which go from $0.6853 + 
0.7283 i$ to $0.9979 + 0.0649 i$ and from $0.6853 - 0.7283 i$ to $0.9979 - 
0.0649 i$ (namely, $\ta$ goes from $0.0648$ to $0.8158$ and from $-0.8158$
to $-0.0648$). The Fourier transforms of the anomalous end modes have peaks at 
$k = 0$ and $\pi$, while the Fourier transform of the end mode with Floquet 
eigenvalue 1 has peaks lying at $k=1.005$ and $2.136$. We also find that the 
expectation value of $\Sigma^y$ is zero in all the end modes, implying
that each mode has equal probabilities of particles and holes. Thus all the 
features for this case are qualitatively similar to the results for 
case (i) that we have presented in Figs.~\ref{fig:ipr} - \ref{fig:fourier2}.
We will see in the next section that various bulk-boundary correspondences 
also work similarly for cases (i) and (ii) except for the winding number.
 
Once again, the anomalous end modes disappear when the chemical 
potential is moved away from zero; for the parameters given in the previous
paragraph, we find that those end modes are no longer present when $|\mu|$
is $0.031$ or larger.

\section{Bulk-boundary correspondence}
\label{sec:bbc}

We will now study if there are any bulk-boundary correspondences which can
help us to understand some of the properties of the end modes discussed
in Sec.~\ref{sec:end}. To this end, we will consider a bulk system with 
periodic boundary conditions. 

As a specific example, we will again consider a periodic driving of the form 
$\ga (t) = \ga_0 [ 1 + a \cos (\om t)]$, with $\ga_0 = 1$, $\De = 0.8$, $\mu 
= 0$, $a=0.5$ and $\om = 1.7$ as in Figs.~\ref{fig:ipr}-\ref{fig:fourier2}. 
With periodic boundary conditions, the momentum $k$ is a good quantum
number; the system therefore decomposes into a sum of subsystems labeled
by $k$. For each value of $k$, we have a Floquet operator which 
is a $2 \times 2$ matrix defined as
\bea U_k &=& {\cal T} e^{-i \int_0^T dt ~h_k (t)}, \non \\
h_k (t) &=& \{2 \ga_0 [1 + a \cos (\om t)] \cos k - \mu \} ~\tau^z + 
2 \De \sin k ~\tau^y. \non \\
&& \label{uk1} \eea
Since each of the terms $e^{-i dt h_k}$ is an SU(2) matrix (a $2 \times 2$
matrix with determinant equal to 1), $U_k$ is also an SU(2) matrix. Further,
the symmetry $h_k (T-t) = h_k (t)$ and the fact that $\tau^z$ ($\tau^y$) is 
a symmetric (antisymmetric) matrix imply that $\tau^z U_k^T \tau^z = U_k$.
(This is similar to the relation given in Eq.~\eqref{trssym} for an open 
chain). This implies that $U_k$ can be written as 
\beq U_k ~=~ e^{i (a_{2,k} \tau^y ~+~ a_{3,k} \tau^z)}, \label{uk2} \eeq
where $(a_{2,k}, a_{3,k})$ can be found uniquely by imposing the condition
$0 < \sqrt{a_{2,k}^2 + a_{3,k}^2} < \pi$. It is possible to impose 
this condition as long as $U_k \ne \pm I_2$. If $U_k = \pm I_2$ for any value
of $k$, the winding number does not exist for the following reason.
We can map the operator $U_k$ in Eq.~\eqref{uk2} to a point on the surface
of a sphere with polar angles $(\al,\be)$, where $\al = \sqrt{a_{2,k}^2 + a_{3,k}^2}$ and $\be = \tan^{-1} (a_{2,k}/a_{3,k})$. We get a closed curve on
the sphere if we take $k$ to go from zero to $2\pi$; the winding number of this
curve is well-defined only if the curve does not pass through either the 
north pole or the south pole (i.e., $\al = 0$ or $\pi$).

\begin{figure}[htb] \ig[width=3.4in]{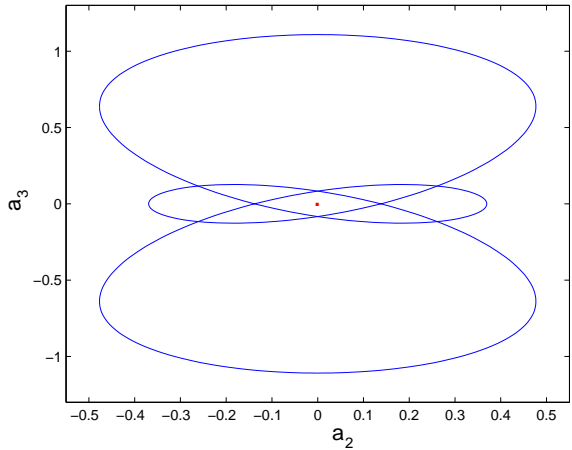}
\caption[]{Closed curves in the $y-z$ plane defined by $(a_{2,k},a_{3,k})$ 
for a system with periodic boundary conditions and the same parameter values 
as in Fig.~\ref{fig:ipr}. The winding number around the origin $(0,0)$ (shown 
by a small red square) is found to be $-3$.} \label{fig:wind} \end{figure}

If we take $(a_{2,k},a_{3,k})$ to define the coordinates of a point in the 
$y-z$ plane, we get a closed curve as $k$ goes from zero to $2\pi$. Figure 
\ref{fig:wind} shows this curve for the parameter values given above. [We
observe that the figure is symmetric under reflection about the line $a_2 = 0$;
this is because of the relations $a_{2,k} = - a_{2,2\pi-k}$ and $a_{3,k} = 
a_{3,2\pi-k}$ which follow from Eq.~\eqref{uk1}. The figure is also symmetric 
under reflection about the line $a_3 = 0$; this is because we have chosen
$\mu = 0$ which implies $a_{2,k} = a_{2,\pi-k}$ and $a_{3,k} = - a_{3,\pi-k}$].
We can then find the values of the two topological invariants defined in 
Sec.~\ref{sec:kit}, namely, the winding number $W$ of the curve around the 
origin $(0,0)$, and $\nu = sgn (a_{3,0} a_{3,\pi})$. For the curve shown in 
Fig.~\ref{fig:wind} we find that $W=-3$ and $\nu = -1$. We see that $|W|$ 
precisely matches the number of modes at each end of the open chain with 
Floquet eigenvalue equal to 1, and the value of $\nu$ correctly indicates 
that the number of end modes with Floquet eigenvalue 1 is odd.

We observe that there are certain values of parameters for which 
$\sqrt{a_{2,k}^2 + a_{3,k}^2}$ is equal to zero or $\pi$, namely, $U_k$ is 
equal to $\pm I_2$ for some value of $k$; the winding number and $\nu$ are 
both undefined in those cases. Looking at Eqs.~\eqref{uk1} and \eqref{uk2}, 
we see that this happens 
for all values of $a$ if $2(\pm 2 \ga_0 - \mu)/\om$ is an integer since $U_0$ 
or $U_\pi$ is then equal to $\pm I_2$. This also happens for all values of 
$a$ if $2 \sqrt{\mu^2 + 4 \De^2}/\om$ is an integer since $U_{\pi/2}$ and 
$U_{3\pi/2}$ are then equal to $\pm I_2$. For the range of parameters used in 
Fig.~\ref{fig:quasi_real}, namely, $\ga_0 = 1, ~\De = 0.8$ and $\mu = 0$, we 
see in that figure that the gap between the bulk Floquet eigenvalues
and the eigenvalue at 1 (i.e., $\ta = 0$) closes at several values of $\om$ 
such as $\om = 2, 1.6, 1, 0.8, 0.667$ and $0.533$; these values agree with the 
conditions on $\om$ given above where the winding number is not defined.

\begin{figure}[htb] \ig[width=3.4in]{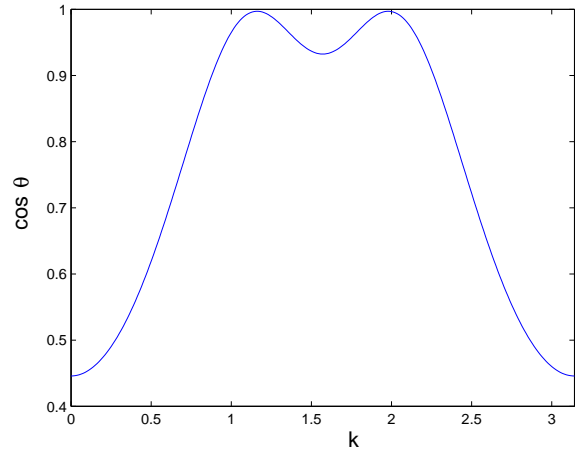}
\caption[]{Real part of Floquet eigenvalue, $\cos \ta_k$, versus $k$ for a 
system with periodic boundary conditions and the same parameter values as in 
Fig.~\ref{fig:ipr}.} \label{fig:real} \end{figure}

Next, we look at the eigenvalues of $U_k = e^{\pm i \ta_k}$, where
$\ta_k = \sqrt{a_{2,k}^2 + a_{3,k}^2}$. 
In Fig.~\ref{fig:real}, we show the real part of the Floquet eigenvalue
(namely, $\cos \ta_k$) versus $k$ for $0 \le k \le \pi$. (The figure looks 
similar for $\pi \le k \le 2\pi$ since $\cos \ta_{2\pi -k} = \cos \ta_k$). 
We see that it has five extrema; these are at $k=0$ and $\pi$ where
$\cos \ta_k = 0.4457$, $k=1.162$ and $1.979$ where $\cos \ta_k = 0.9972$, 
and $k = \pi/2$ where $\cos \ta_k = 0.9325$. [The value of $\cos \ta_k$
at $k=0$ and $\pi$ can be obtained easily from Eq.~\eqref{uk1} since the 
coefficient of $\tau^y$ is then zero. We find that $\cos \ta_k = \cos [ 
(2\ga_0 - \mu) T]$, independent of the values of $\De$ and $a$].

We note that the range of values of the Floquet eigenvalues for the bulk system
with periodic boundary conditions precisely matches the range of values of 
the continuous band of Floquet eigenvalues for the open chain as shown in 
Fig.~\ref{fig:floeig}; this could have been anticipated. However, we now make 
an additional observation. Namely, some (but not all) of the extrema of the 
Floquet eigenvalues of the bulk system have a close correspondence with the 
anomalous end modes of the open chain in two different ways. First, the values 
of the bulk Floquet eigenvalues at the extrema at $k=0$ and $\pi$ are close 
to the Floquet eigenvalues $0.4333 \pm 0.9012 i$ for four of the end modes of 
the open chain. Second, the peaks of the Fourier transforms of these end modes
lie at $k=0$ and $\pi$. Interestingly, we also see that the values of the bulk 
Floquet eigenvalues at $k=1.162$ and $1.979$ are close to the Floquet 
eigenvalue of 1 for six of the end modes of the open chain, and the peaks of 
the Fourier transforms of these end modes lie at $k=1.162$ and $1.979$.

It would be useful to understand why there is such a correspondence between
the extrema of the Floquet eigenvalues of the bulk system and the end modes 
of an open chain. We offer a speculation here. The fact of an extremum of 
the bulk Floquet eigenvalues near a particular value, say, $e^{i \ta'}$
at $k= k'$, means that the density of states $\rho (\ta) \equiv \int dk ~\de 
(\ta - \ta_k)$ diverges as we approach $\ta = \ta'$. The presence of a large 
number of bulk states near $(k',\ta')$ perhaps makes it easy for an open chain 
to superpose those states to form modes which are localized at the ends.
Such end modes will then naturally have a Floquet eigenvalue close to
$e^{i \ta'}$ and a Fourier transform whose peak is close to $k'$.

We will now comment briefly about the example of case (ii), $\ga (t) = \ga_0 
e^{i a \cos (\om t)}$, with $\ga_0 = 1, ~\De = 0.8, ~\mu = 0, ~\om = 1.7$ and 
$a=0.4$ which was discussed at the end of Sec.~\ref{sec:kit}. 
In the bulk of this system,
the Floquet operator for momentum $k$ is a $2 \times 2$ matrix given by 
\bea U_k &=& {\cal T} e^{-i \int_0^T dt ~h_k (t)}, \non \\
h_k (t) &=& \{2 \ga_0 \cos (a \cos (\om t)) \cos k - \mu \} ~\tau^z \non \\
&& +~ 2 \ga_0 \sin (a \cos (\om t)) \sin k ~I_2 ~+~ 2\De \sin k ~\tau^y. 
\non \\
&& \label{uk3} \eea
Using arguments similar to those presented after Eq.~\eqref{uk1}, the
facts that $h_k (T-t) = h_k (t)$ and $I_2$ commutes with both $\tau^y$ and
$\tau^z$ implies that Eq.~\eqref{uk3} can be written as
\beq U_k ~=~ e^{i (a_{0,k} I_2 ~+~ a_{2,k} \tau^y ~+~ a_{3,k} \tau^z)}. 
\label{uk4} \eeq
If $(a_{0,k},a_{2,k},a_{3,k})$ were all non-zero, they would define a closed 
curve in three dimensions as $k$ goes from zero to $2\pi$, and it would not be
possible to define a winding number. However, in this problem we find that
\beq a_{0,k} ~=~ -~ \int_0^T ~dt ~2\ga_0 \sin (a \cos( \om t)) \sin k \eeq 
is equal to zero 
for all $k$ because the integrand changes sign under the shift $t \to t + 
\pi/\om$. Hence, we only have two non-zero variables $(a_{2,k},a_{3,k})$ as
in Eq.~\eqref{uk2} and we can therefore define a winding number $W$. For
the parameters given above, we find that $W=3$ which does {\it not} match 
the number of modes (namely, one) at each end of the open chain with Floquet 
eigenvalue 1. However, we find that $\nu = -1$ which agrees with the fact 
that the number of end modes with Floquet eigenvalue 1 is odd.

We now make some comments about the anomalous end modes whose Fourier 
transforms have peaks at $k=0$ and $\pi$, and we present a qualitative argument
about why these modes disappear when $\mu$ moves away from zero. For $k=0$ 
and $\pi$, it is easy to compute the Floquet eigenvalues of the bulk system.
For case (i), $\ga(t) = \ga_0 [1 + a \cos (\om t)]$, we find from 
Eq.~\eqref{uk1} that 
\bea U_0 &=& \exp [-i (2\pi/\om) (2\ga_0 - \mu) \tau^z], \non \\
U_\pi &=& \exp [-i (2\pi/\om) (-2\ga_0 - \mu) \tau^z]. \eea
If $\mu = 0$, the eigenvalues of $U_0$ and $U_\pi$ are both equal to
$e^{\pm i (4 \pi/\om) \ga_0}$. In an open chain, the breaking of translation 
invariance and the degeneracy of the eigenvalues of $U_0$ and $U_\pi$ 
means that the eigenvectors of $U_0$ and $U_\pi$ can easily hybridize.
This is probably the reason why they can combine to form end modes,
whose Floquet eigenvalues lie close to $e^{\pm i (4 \pi/\om) \ga_0}$ if
the driving amplitude $a$ is small. A similar argument works for case (ii),
$\ga (t) = \ga_0 e^{i a \cos (\om t)}$, where Eq.~\eqref{uk3} implies that
\bea U_0 &=& \exp [-i (2\pi/\om) (2\ga_0 J_0 (a) - \mu) \tau^z], \non \\
U_\pi &=& \exp [-i (2\pi/\om) (-2\ga_0 J_0 (a)- \mu) \tau^z], \eea
where we have used the identity~\cite{abram}
\beq \frac{1}{2\pi} ~\int_0^{2\pi} d\ta ~\cos (a \cos \ta) ~=~ J_0 (a). \eeq
Once again, if $\mu = 0$, the eigenvalues of $U_0$ and $U_\pi$ are degenerate
and equal 
to $e^{\pm i (4 \pi /\om) \ga_0 J_0 (a)}$, and the corresponding eigenvectors
can hybridize and form end modes for an open chain. However, when $\mu$ is
moved away from zero, the eigenvalues of $U_0$ and $U_\pi$ no longer remain 
degenerate; this may make it difficult for their eigenvectors to 
hybridize and form end modes.

We end this section by pointing out some other numerical observations. All 
the results presented above were for cases where the frequency is of the order
of or larger than the other energy scales of the system such as $\ga_0$ and
$\De$ and when the driving amplitude is less than or of order 1. When we move 
away from this regime, the results can change as follows. First, if the 
amplitude is much larger than 1, we sometimes find end modes whose Floquet 
eigenvalues lie far outside the range of the Floquet eigenvalues of the bulk 
modes. Then the Floquet eigenvalues of those end modes do not lie close 
to any of the extrema of the Floquet eigenvalues of the bulk modes; hence this
aspect of the bulk-boundary correspondence does not work for such end modes.
Second, for frequencies much smaller than $\ga_0$ and $\De$, we sometimes 
find end modes whose Floquet eigenvalues lie within the range of the Floquet 
eigenvalues of the bulk modes. We can identify such end modes in the plot of 
the IPRs as in Fig.~\ref{fig:ipr} but not in a plot of the Floquet 
eigenvalues as in Fig.~\ref{fig:floeig}.

\section{Floquet-Magnus expansion}
\label{sec:mag}

In this section, we will study the properties of the system when the driving
frequency $\om$ is much larger than the hopping $\ga_0$ and the 
superconducting pairing $\De$. In this limit we can use a Floquet-Magnus
expansion in powers of $1/\om$ to find an effective 
Hamiltonian~\cite{mikami,bukov}. 

Suppose that a Hamiltonian $H$ varies in time with a period $T=2\pi/\om$. 
Then we can write
\bea H &=& \sum_{n=-\infty}^\infty ~H_n ~e^{-in\om t}, \non \\
{\rm where} ~~~~H_n &=& \frac{1}{T} ~\int_0^T ~dt ~H(t) ~e^{in\om t}. 
\label{ht} \eea
We now write the Floquet operator $U ~=~ {\cal T} \exp (-i \int_0^T dt H)$ as 
\beq U ~=~ e^{-i H_{eff} T}, \label{heff1} \eeq
where $H_{eff}$ is called the effective Floquet Hamiltonian.
Then the first three terms in the Floquet-Magnus expansion are given 
by~\cite{mikami,bukov}
\bea H_{eff} &=& H_0 ~+~ \sum_{n \ne 0} \frac{[H_{-n}, H_n]}{2n \om} 
~+~ \sum_{n \ne 0} \frac{[[H_{-n}, H_0], H_n]}{2 n^2 \om^2} \non \\
&& +~ \sum_{m \ne 0} ~\sum_{n \ne 0,m} \frac{[[H_{-m}, H_{m-n}], H_n]}{3 m n 
\om^2}. \label{heff2} \eea

For a system with periodic boundary conditions, we can calculate the 
effective Hamiltonian in momentum space. We will use the symbols $h_{n,k}$ to 
denote the Fourier components of $h_k (t)$ as defined in Eq.~\eqref{ht} and 
$h_{eff,k}$ to denote the effective Hamiltonian. In the case that the hopping 
is periodically driven as $\ga (t) = \ga_0 [ 1 + a \cos (\om t)]$, we see from
Eq.~\eqref{uk1} that the only non-zero components $h_{n,k}$ are given by
\bea h_{0,k} &=& [2\ga_0 \cos k - \mu] ~\tau^z + 2 \De \sin k ~\tau^y, \non \\
h_{1,k} &=& h_{-1,k} ~=~ a \ga_0 \cos k ~\tau^z. \label{hnk1} \eea
This implies that the term of order $1/\om$ in $h_{eff,k}$ vanishes
since $[h_{1,k}, h_{-1,k}] = 0$. To order $1/\om^2$, we obtain
\bea && \frac{[[h_{-1,k},h_{0,k}],h_{1,k}]}{2\om^2} ~+~ \frac{[[h_{1,k},
h_{0,k}], h_{-1,k}]}{2\om^2} \non \\
&=& - ~\frac{8 a^2 \ga_0^2 \De}{\om^2} ~\cos^2 k \sin k ~\tau^y. \eea
Using Eq.~\eqref{heff2}, we find that
\bea h_{eff,k} &=& (2\ga_0 \cos k - \mu) ~\tau^z \non \\
&& + 2 \De \sin k ~(1 ~-~ \frac{4 a^2 \ga_0^2}{\om^2} \cos^2 k)~\tau^y. 
\label{heff3} \eea

For the parameters used in Fig.~\ref{fig:quasi_real},
$\ga_0 = 1, ~\De = 0.8$ and $\mu = 0$,
we find that the Floquet operator $U_k = \exp (-i h_{eff,k} T)$ is not
equal to $\pm I_2$ for any value of $k$ if $\om > 4$ and $4 a^2/\om^2 < 1$.
We then find that the winding number is $-1$ and the other topological 
invariant $\nu = sgn (a_{3,0} a_{3,\pi}) = -1$. We therefore expect that 
each end of an open chain will have one end mode with Floquet eigenvalue
equal to 1 (i.e., $\ta = 0$) if $\om$ is large enough. This agrees with what 
we observe in Fig.~\ref{fig:quasi_real}.

We can similarly analyze the case where $\ga (t) = \ga_0 e^{ia \cos (\om t)}$.
This turns out to be more complicated than the previous case in that $h_{n,k}$
is now non-zero for all values of $n$. Namely, using the 
identities~\cite{abram}
\bea \cos [a \cos (\om t)] &=& J_0 (a) + 2 \sum_{m=1}^\infty (-1)^m 
J_{2m} (a) \cos (2m \om t), \non \\
\sin [a \cos (\om t)] &=& - 2 \sum_{m=1}^\infty (-1)^m J_{2m-1} (a) 
\cos [(2m-1) \om t], \non \\
&& \eea
we find that
\bea h_{0,k} &=& [2\ga_0 J_0 (a) \cos k - \mu] ~\tau^z + 2 \De \sin k ~\tau^y,
\non \\
h_{2m,k} &=& h_{-2m,k} ~=~ 2 (-1)^m \ga_0 \cos k ~J_{2m} (a) ~\tau^z, \non \\ 
h_{2m-1,k} &=& h_{-2m+1,k} ~= - 2 (-1)^m \ga_0 \sin k ~J_{2m-1} (a) ~I_2, 
\non \\
&& \label{hnk2} \eea 
where $m=1,2,3, \cdots$ in the last two lines of Eqs.~\eqref{hnk2}. We see 
that $h_{n,k} = h_{-n,k}$ for all $n$; hence the terms of order $1/\om$ 
in $h_{eff,k}$ again vanish, and we have to go to order $1/\om^2$. We will 
not discuss this case further.

\section{Conclusions}
\label{sec:con}

In this paper we have numerically shown that periodic driving of either the 
magnitude or the phase of the nearest-neighbor hopping amplitude in a 
one-dimensional model of electrons with $p$-wave superconductivity can 
generate end modes in a long and open chain. The end modes are of two types: 
some have Floquet eigenvalues $e^{i\ta}$ equal to $\pm 1$ (i.e., $\ta = 0$
or $\pm \pi$), while the others, called anomalous, have Floquet eigenvalues 
different from $\pm 1$ in complex conjugate pairs (i.e., $\ta \ne 0$ or 
$\pm \pi$). We observe that the anomalous end modes disappear if the 
chemical potential is moved sufficiently away from zero. We also find that 
a sufficiently large driving amplitude is required to produce the anomalous 
end modes. To the best of our knowledge, the 
anomalous end modes have not been seen earlier in models in one dimension; 
see, for instance, Ref.~\onlinecite{jiang,liu,thakur1,asboth} where all the 
end modes are found to have Floquet eigenvalues equal to $\pm 1$.

We have studied if there are any bulk-boundary correspondences and any
topological invariants which can relate the bulk system with periodic 
boundary conditions and the end modes of an open chain. We find that the 
$Z$-valued winding number of the Floquet operator in the bulk matches the 
number of modes at each end of a chain with Floquet eigenvalue equal to 1 
if the magnitude of the hopping is periodically driven but not if the phase 
of the hopping is driven. In contrast to this, there is a $Z_2$-valued 
topological invariant which always agrees with the number of modes at each 
end with Floquet eigenvalue equal to 1 being an even or odd integer. 
This is in agreement with the discussion of time-reversal symmetry and
topological invariants in Sec.~\ref{sec:kit}: the case in which the magnitude 
of the hopping is periodically driven is time-reversal symmetric and allows
a $Z$-valued topological invariant, while the case in which the phase of the 
hopping is periodically driven is not time-reversal symmetric and 
only allows a $Z_2$-valued topological invariant. However, the anomalous
end modes with Floquet eigenvalues different from $\pm 1$ appear whether or
not time-reversal symmetry is broken, and there does not seem to be any 
topological invariant (or any simple function of the different
parameters of the system) which matches the number of such end modes.
There is however an interesting bulk-boundary correspondence if the driving 
amplitude is not too large. We find that 
the Floquet eigenvalues of the anomalous end modes lie close to the ends of
the band of Floquet eigenvalues of the bulk system (which are labeled by a 
momentum $k$). Further, the Fourier transforms of the wave functions of these 
end modes have peaks at values of $k$ which match closely with the values of 
$k$ where the bulk bands of Floquet eigenvalues have extrema. For instance,
the Fourier transforms of the anomalous end modes have peaks at $k=0$ and 
$\pi$ where the Floquet eigenvalues of the bulk system have extrema. While
we have presented some qualitative arguments, it would be useful to completely
understand the reasons for these correspondences between the end modes and 
the bulk system.

We have used a Floquet-Magnus expansion to find the effective Floquet 
Hamiltonian in the limit that the driving frequency is much larger than 
the other energy scales of the system, namely, the hopping and superconducting 
pairing. We have found that in this limit, the number of end modes is the 
same as that found when there is no driving.

There has been much excitement in recent years about the possibility of 
detecting Majorana modes in time-independent systems of superconducting 
nanowires~\cite{kouwenhoven,deng,rokhinson,das,finck} following some 
theoretical proposals~\cite{lutchyn1,oreg,alicea,stanescu}. A zero bias peak 
has been observed in the tunneling conductance into one end of the nanowire, 
and it has been suggested that this is the signature of a Majorana end mode. 
Our results can be tested in similar systems by applying an oscillating
electric field (such as electromagnetic radiation) to the nanowire. One can 
study if the presence of the end modes produced by periodic driving modifies 
the sub-gap conductance peaks in some way; this has 
been studied in a related model in Ref.~\onlinecite{kundu}. We have shown 
in this paper that there are two kinds of end modes: Majorana end modes 
with Floquet eigenvalues equal to $\pm 1$ and anomalous end modes with Floquet 
eigenvalues which differ from $\pm 1$; it would be useful to know if these 
contribute to conductance peaks in different ways. A question which needs 
to be examined in this context is how the end modes appear in the steady 
state of the system after the oscillatory electric field is switched on.
This would require an analysis of various relaxation mechanisms which
may be present in the system~\cite{lind1}. Finally, it would be interesting
to study how the different end modes are affected by disorder in, say, 
the chemical potential~\cite{motrunich,brouwer1,lobos,cook,pedrocchi}.

\section*{Acknowledgments}

S.S. and S.N.S. thank Department of Science and Technology, India for KVPY 
Fellowships. D.S. thanks Department of Science and Technology, India for 
Project No. SR/S2/JCB-44/2010 for financial support.

\end{document}